\documentclass[fleqn,usenatbib]{mnras}

\usepackage{newtxtext,newtxmath}
\usepackage[T1]{fontenc}

\DeclareRobustCommand{\VAN}[3]{#2}
\let\VANthebibliography\thebibliography
\def\thebibliography{\DeclareRobustCommand{\VAN}[3]{##3}\VANthebibliography}

\usepackage{graphicx}	% Including figure files
\usepackage{amsmath}	% Advanced maths commands
\usepackage{tablefootnote}
\title[Linking dust geometry and SFR]{EMU/GAMA: Understanding the link between dust geometry and star formation rate in galaxies}

\author[U. T. Ahmed et al.]{
U. T. Ahmed,$^{1}$\thanks{E-mail: ummeetania.ahmed@mq.edu.au}
A. M. Hopkins,$^{1}$
B. Farley,$^{2}$
M. Cowley,$^{2,3}$
S. Phillipps,$^{4}$
Y. A. Gordon,$^{5}$
B. W. Holwerda,$^{6}$
\newauthor
D. J. Bomans,$^{7}$
K. A. Pimbblet,$^{8}$
S. Casura,$^{9}$
A. Battisti,$^{10,11,12}$
D. A. Leahy,$^{13}$
and J. Prathap$^{1}$
\\
$^{1}$School of Mathematical and Physical Sciences, 12 Wally’s Walk, Macquarie University, NSW 2109, Australia\\
$^{2}$School of Chemistry \& Physics, Faculty of Science, Queensland University of Technology, Brisbane, QLD 4000, Australia\\
$^{3}$University of Southern Queensland, Centre for Astrophysics, West Street, Toowoomba, QLD 4350 Australia\\
$^{4}$Astrophysics Group, School of Physics, University of Bristol, Tyndall Ave,, Bristol BS8 1TL, UK\\
$^{5}$ Department of Physics, University of Wisconsin-Madison, Madison, WI 53706-1390, USA\\
$^{6}$Department of Physics and Astronomy, University of Louisville, Natural Science Building 102, Louisville, KY 40292, USA\\
$^{7}$Ruhr-Universität Bochum, Astronomisches Institut, Universitätsstr 150 / GAFO03, 44801 Bochum, Germany\\
$^{8}$Centre of Excellence for Data Science, AI, and Modelling (DAIM), University of Hull, Cottingham Road, Kingston-upon-Hull, HU6 7RX, UK\\
$^{9}$Hamburger Sternwarte, Universität Hamburg, Gojenbergsweg 112, 21029 Hamburg, Germany\\
$^{10}$International Centre for Radio Astronomy Research, University of Western Australia, 7 Fairway, Crawley, WA 6009, Australia\\
$^{11}$Research School of Astronomy \& Astrophysics, The Australian National University, Mt Stromlo Observatory D.115, Cotter Rd, Weston Creek, ACT 2611, Australia\\
$^{12}$ARC Centre of Excellence for All Sky Astrophysics in 3 Dimensions (ASTRO-3D), Australia\\
$^{13}$Department of Physics and Astronomy, University of Calgary, Calgary, Alberta, T2N 1N4, Canada\\
}
\date{Accepted XXX. Received YYY; in original form ZZZ}

\pubyear{\the\year{}}

\begin{document}
\label{firstpage}
\pagerange{\pageref{firstpage}--\pageref{lastpage}}
\maketitle

\begin{abstract}
We investigate the link between dust obscuration and star formation rate (SFR) estimators using H$\alpha$ luminosities, $L_{{\rm H}\alpha}$ and radio luminosities, $L_{\rm radio}$. 
While the Balmer Decrement (BD) is the standard tool for attenuating H$\alpha$, it can fail to account for optically thick dust regions from which H$\alpha$ cannot escape. The ratio of SFRs estimated from radio luminosity and $L_{{\rm H}\alpha}$ corrected for dust attenuation using the BD can be used as a probe of the degree of optically thick dust affecting H$\alpha$. We explore this combining Evolutionary Map of
the Universe (EMU) Early Science data at $887.5\,$MHz in the Galaxy and Mass Assembly (GAMA) G23 field with the GAMA data. We draw on dust estimates using both the BD and the dust mass, $M_d$, demonstrating systematic underlying dependencies with $L_{\rm radio}/L_{{\rm H}\alpha}$. 
By combining the BD and $M_d$, we quantify how they relate to $L_{\rm radio}/L_{{\rm H}\alpha}$ and other galaxy properties. 
We use this approach to establish an H$\alpha$ SFR ($SFR_{{\rm H}\alpha}$) estimator that accounts for optically thick dust, demonstrating its consistency with conventional SFR indicators.
Rather than proposing a replacement for standard correction methods, we aim to demonstrate the effectiveness of a dust geometry model that integrates both BD and $M_d$ parameters. 
This $H_{dust}$ parameter provides a more physically complete representation of optically thick environments that traditional methods may overlook, and can produce SFR estimates consistent with traditional techniques.

\end{abstract}


%%%%%%%%%%%%%%%%%%%%%%%%%%%%%%%%%%%%%%%%%%%%%%%%%%

%%%%%%%%%%%%%%%%% BODY OF PAPER %%%%%%%%%%%%%%%%%%

\section{Introduction}
\label{sec:int}

The star formation rate (SFR) is a fundamental parameter in understanding galaxy formation and evolution. Extensive work has established numerous approaches for estimating the SFR through diverse observational methods. These include the analysis of nebular emission lines \citep[e.g.,][]{Glazebrook_2004, Westra_2010}, which trace the presence of massive stars \citep{2013MNRAS.434..209B, 2016MNRAS.462...92J} as well as investigations into ultraviolet (UV), and infrared (IR) emission \citep[e.g.,][]{2005ApJ...619L..47S, 2008ApJS..175...48R, 2025arXiv250319843L} sensitive to the direct emission from young star forming regions and from emission reradiated by dust, respectively. Additionally, the SFR can be probed through radio emission originating in supernova remnants \citep[e.g.,][]{2008MNRAS.386.1695S}, X-ray emission \citep[e.g.,][]{2003MNRAS.345..939G, 2003MNRAS.339..793G} produced by high mass X-ray binaries, and gamma-ray bursts \citep[e.g.,][]{2009ApJ...705L.104K} stemming from massive stellar explosions.

Galaxy spectral energy distributions (SEDs) encapsulate the physical properties of galaxies, such as their SFR, stellar mass, $M_*$, and dust mass, $M_d$. SED fitting is a standard technique to measure these galaxy properties \citep[e.g.,][]{2008MNRAS.388.1595D, 2020MNRAS.495..905R, 2020MNRAS.498.5581B, Pacifici_2023}.
The Balmer decrement (BD) provides a measure of dust attenuation in the HII regions such that higher ratio indicates more attenuation \citep{2010MNRAS.409..421G, 2012MNRAS.426..892G, 2013ApJ...763..145D, 2015ApJ...806..259R, 2016ApJ...817L...9N, 2020MNRAS.495.2305G}. Conversely, $M_d$ is typically estimated through sub-mm/IR measurements \citep{2010A&A...518L.154S, 2011ApJ...738...89S, 2012ApJ...748..123S, 2017MNRAS.465...54C, 2022ApJ...928...68S} or modelling the galaxy’s SED \citep{2008MNRAS.388.1595D, 2020MNRAS.495..905R}. 
The cold dust population constitutes the bulk of $M_d$ in galaxies, shaping the SED particularly at far-infrared (FIR)/sub-mm wavelengths \citep{2000ApJ...528..799W, 2001ApJ...551..277M}. Unlike the BD, SED fitting incorporates various parameters, across different wavelengths, that influence how dust attenuates the stellar light emitted by galaxies. These dust related parameters are fitted, together with other stellar population parameters, to find the best fit model that reproduces the observed SED of the galaxy.

In the absence of IR constraints for galaxies, obtaining robust SED parameters requires navigating the inherent degeneracies between stellar age, dust attenuation, and star formation history. A primary alternative is the IRX-$\beta$ relation \citep[e.g.,][]{Meurer1999, Reddy2012}, though this method remains sensitive to the choice of attenuation law \citep[e.g.,][]{2000ApJ...533..682C,calzetti1994dust}. Recently, \citet[][and references therein]{2024ApJ...962...59O} demonstrated that even without IR data, reliable recovery of parameters such as total IR luminosity and dust corrected SFRs is possible by optimising SED fitting strategies against known ground truth samples from deep surveys. To mitigate uncertainties, studies increasingly employ SED tools such as \textsc{ProSpect} \citep{2020MNRAS.495..905R, 2020MNRAS.498.5581B}, MAGPHYS \citep{2008MNRAS.388.1595D} or CIGALE \citep{2019A&A...622A.103B}, which physically link the energy absorbed in the UV/optical to the re-emitted dust luminosity. Furthermore, the use of empirical proxies, such as the FIR-radio correlation \citep[e.g.,][]{2017A&A...602A...4D} allows for a statistical grounding of SFR across broad populations.

\citet{2025PASA...42...67F} presented a new method to quantify dust in galaxies by combining the BD and $M_d$. They introduced two new parameters that link BD and $M_d$ using SFR across multiple wavelengths, including UV, H$\alpha$, and FIR. These novel metrics are sensitive to optically thick dust that affects short wavelength emission. \citet{2025PASA...42...67F} examined how these new ``dust geometry'' parameters vary with galaxy mass, SFR, and specific SFR. Their results showed that they effectively trace dust geometry and support the ``maximal foreground screen'' model in starburst galaxies.
Continuing from that analysis, the work presented here explicitly links these two tracers of dust properties in galaxies, BD and $M_d$, to address discrepancies between SFR indicators. Motivated by the early work of \citet{2001AJ....122..288H}, we apply this method to minimise the discrepancy between the SFR measurements from $L_{{\rm H}\alpha}$ and $L_{\rm radio}$. This approach enables us to demonstrate H$\alpha$ based SFR measurements by accounting for emission that is otherwise obscured by optically thick dust,
thus providing star formation rates that remain consistent with established indicators.

Many studies have made comparisons between different SFR indicators at different wavelengths, and established detailed relations in between those indicators \citep[e.g.,][]{2001AJ....122..288H, 2001ApJ...558...72S, 2002A&A...383..801B, 2003ApJ...586..794B, 2016MNRAS.461..458D}. For a comprehensive understanding of the star formation history (SFH) of the universe, it is typical to rely on estimates of the SFR in galaxies arising from different methods, including commonly from radio, IR, optical, and UV \citep[see, e.g., the review by][and references therein]{1992ARA&A..30..575C, 2012ARA&A..50..531K}.  
These methods arise from different physical processes, which can lead to variations in the derived SFR values. UV and optical wavelengths are heavily attenuated by dust, leading to under-estimation of SFR in highly obscured regions. In contrast, IR and radio wavelengths are less affected by dust obscuration and may provide more reliable estimates in dusty environments. 
Variations in stellar populations, dust properties, and metallicity can introduce systematic biases that complicate comparisons between SFR estimates derived from these various methods. Individual tracers are sensitive to the chosen stellar initial mass function (IMF) and UV and H$\alpha$ depend on differential attenuation between young birth clouds and the diffuse interstellar medium (ISM) \citep[e.g.,][]{2000ApJ...539..718C}. Multi-wavelength approaches, including SED fitting, that incorporate radio and IR data provide a more robust approach by accounting for dust obscured star formation and reducing systematic uncertainties associated with single component attenuation models.

The current study explores the relationship between the radio SFR and H$\alpha$ SFR ($SFR_{{\rm H}\alpha}$) by incorporating additional dust properties. Radio luminosities ($L_{\rm radio}$) provide a measure of SF unaffected by dust, presuming that active galactic nuclei (AGN) emission has been excluded. $L_{{\rm H}\alpha}$, which is sensitive to dust obscuration, offers complementary insights upon correction for dust effects.
In order to compare $L_{{\rm H}\alpha}$ to $L_{\rm radio}$, we need to make dust corrections to the H$\alpha$ measurements. Such dust corrections are well-established, and many authors have introduced different dust models to accurately account for dust obscuration. For example, the influential early work of \citet{calzetti1994dust}, \citet{ 2000ApJ...533..682C}, and \citet{2001PASP..113.1449C} \citep[see also, e.g.,][]{1984ApJ...287..228N, 1989MNRAS.239..939D, 1992ApJ...393..611W} defines and explores conceptual models for the geometric distribution of dust in galaxies. Moreover, \citet{1989MNRAS.239..939D} concluded that incorrect evaluations of luminosities and SFR may result from an insufficient consideration of dust geometry. 
Rather than replacing standard correction techniques, we show how a dust geometry model can be validated by combining both BD and $M_d$ parameters. By linking these two values, we can better account for the ways dust is distributed.
While traditional Balmer based corrections are widely used, they often face limitations in the most obscured regions of a galaxy. We show that the $H_{dust}$ parameter provides a more physically complete representation of these optically thick environments that traditional methods may overlook. Our approach can be used to make reliable SFR corrections, acting as a complementary tool to established empirical relations. This method is a novel way to represent an optically thick dust measure, consistent with other approaches.

The structure of the paper is as follows. 
We provide a brief introduction to the dataset from the GAMA and EMU surveys, and describe the sample selection criteria in \S~\ref{sec:data}. The analysis of the H$\alpha$ - radio SFR correlation is presented in \S~\ref{subsec:SFR_Halpha_rad}, and the BD dependencies on luminosity ratios are discussed in \S~\ref{subsec:BD_Lrad}.
In \S~\ref{sec:discussion}, we detail our key findings on the link between these new dust parameters and SFR estimates, and the relation between dust geometry and SFR in the radio detected subsample. We present our conclusions in \S~\ref{sec:concl}. Throughout we assume cosmological parameters of $H_0=70\,$km\,s$^{-1}$\,Mpc$^{-1}$, $\Omega_M=0.3$,
$\Omega_\Lambda=0.7$ and $\Omega_{\rm{k}} = 0$. Spectral index $\alpha$ is defined such that flux density $S$ scales with frequency $\nu$ in the sense $S\propto\nu^{\alpha}$.

\section{Data}
\label{sec:data}

In this analysis we begin with the same dataset as in \citet{2024PASA...41...21A}, and briefly reiterate the details here.
We use photometric and spectroscopic data from two main surveys to identify the host galaxies of radio detected SFGs. The multiwavelength catalogue has been created starting with data from the Galaxy and Mass Assembly survey \citep[GAMA;][]{2011MNRAS.413..971D, 2020MNRAS.496.3235B, 2022MNRAS.513..439D}. This is combined with the Kilo-Degree survey \citep[KiDS;][]{2015A&A...582A..62D} and the VISTA\footnote{Visible Infrared Survey Telescope for Astronomy} Kilo-degree INfrared Galaxy survey \citep[VIKING;][]{2012sngi.confE..40S} datasets. The resulting matched dataset is referred to as the GAMA-KiDS-VIKING catalogue \citep[GKV;][]{2020MNRAS.496.3235B}. This has been combined with radio data from the Evolutionary Map of the Universe survey \citep[EMU;][]{2011PASA...28..215N, 2021PASA...38...46N, 2025PASA...42...71H}. The EMU data used here is from early science observations in the GAMA G23 field \citep{2022MNRAS.512.6104G}, at 887.5\,MHz. We use the latest radio observations of G23 \citep{2022MNRAS.512.6104G}, which replace earlier data \citep{2019PASA...36...24L} obtained using the Australian SKA Pathfinder (ASKAP) telescope \citep{2021PASA...38....9H}. For our radio-IR correlation (see \S~\ref{subsec:sample}, for details), we use IR data from the ALLWISE catalogue \citep{2012yCat.2311....0C} for the GKV sample, as identified by \citet{2020ApJ...903...91Y}.

\begin{table}
\centering
\caption{The number of galaxies in GKV and GKV - Radio from our cross-matched optical parent, and radio samples.}
\label{tab:sample}
\begin{tabular}{lcccr} % five columns, alignment for each
\hline
\hline
Catalogue & Number of Sources\\
\hline
GKV & 41\,115\\
GKV ($nQ$\tablefootnote{Redshift quality flag} $\geq$ {\bf 3}) & 40\,088\\
GKV (gold) & 5154\\
GKV (gold) ($M_d$ $S/N \geq 3$) & 1077\\
GKV -- Radio & 7123\\
GKV -- Radio ($nQ$ $\geq$ {\bf 3}) & 6999\\
GKV -- Radio (gold) & 1700\\
GKV -- Radio (gold) ($M_d$ $S/N \geq 3$) & 688\\
\hline
\end{tabular}
\end{table}

\subsection{Galaxy and Mass Assembly}
\label{subsec:gama}

GAMA \citep{2011MNRAS.413..971D, 2022MNRAS.513..439D} is a wide-field, multi-wavelength survey, achieving 98.5\% redshift completeness independent of the target density \citep{2015MNRAS.452.2087L}. The GAMA survey consists of 300\,000 galaxy spectra  across five regions of the sky (G02, G09, G12, G15, and G23) covering a total area of 286 deg$^2$ \citep{2010MNRAS.404...86B, 2010PASA...27...76R, 2011MNRAS.413..971D, 2013MNRAS.430.2047H,2015MNRAS.452.2087L, 2018MNRAS.474.3875B}. This study analyses the galaxies within just one of the fields, G23 with a redshift completeness of 94.2\% \citep{2015MNRAS.452.2087L}. The G23  sample has a limiting magnitude of $i<19.2$ \citep{2015MNRAS.452.2087L}. This field was selected as it has been observed as part of the ASKAP commissioning \citep{2019PASA...36...24L, 2021PASA...38....9H} and again later for EMU early science  \citep{2022MNRAS.512.6104G}.

We use the GAMA DR4 database \citep{2022MNRAS.513..439D}, which includes photometry and measurements following the methodology outlined by \citet{2024PASA...41...21A}. This database contains the \-\verb|gkvScience-v02| (GKV) catalogue, which is the standard UV–IR dataset for GAMA DR4 and we use its precomputed photometry-to- spectroscopy matches. Accurate emission line measurements are needed to select our SFG sample, so we adopt optical line fluxes from \-\verb|GaussFitComplexv05| within the \-\verb|SpecLineSFR| data management unit \citep[DMU;][]{2017MNRAS.465.2671G}. This DMU provides line fluxes, equivalent widths, and redshifts, the latter sourced from \-\verb|SpecAllv27| within the \-\verb|SpecCat| DMU \citep{2015MNRAS.452.2087L}.
We chose to use the measurements from \-\verb|GaussFitComplexv05| instead of \-\verb|GaussFitSimplev05| as this catalogue employs multiple Gaussian fits to enhance accuracy. Nevertheless, \citet{2024PASA...41...21A} demonstrated that simpler fitting approaches produce quantitatively similar outcomes. At the spectral resolution of GAMA \citep[$R\sim 1300$]{2013MNRAS.430.2047H}, H$\alpha$ is not blended with the [N{\sc ii}] lines, and the multiple Gaussian fits of \-\verb|GaussFitComplexv05| ensure that the lines are separately and accurately measured.

Of the 41\,115 galaxies from the G23 field in the \verb|gkvScience-v02| catalogue, there are 40\,088 galaxies that meet the redshift quality flag, $nQ \ge 3$ high quality redshift standard, forming our ``GKV ($nQ \ge 3$)'' sub-sample (Table~\ref{tab:sample}). We further limit our sample by imposing a series of quality restrictions. We require that H$\alpha$ and H$\beta$ are in emission and that we have detections of the [O{\sc iii}] and [N{\sc ii}] lines. We also require the $\text{IS\_BEST}$ flag to be true, to select the highest S/N spectrum if multiple spectra are available. This leaves 17\,547 sources. 
Next, we select only sources with line fitting flags ${\rm NPEG}=0$ for both H$\alpha$ and H$\beta$, ensuring that none of the fitted parameters were forced to the limits of their allowed ranges. This results in a sample of 9\,774 sources. Finally, we impose signal-to-noise thresholds on the Balmer lines, $S/N$ (H$\alpha$) $\geq 3$, and $S/N$ (H$\beta$) $\geq 3$.
Given these $S/N$ thresholds, we have very few low EW objects in our sample. This means we can be confident that any H$\alpha$ derived SFRs are not erroneously dominated by diffuse ionised gas, which may be the case for low EW emission line objects \citep[e.g.,][]{2010MNRAS.403.1036C}.
This finally produces the 5\,154 galaxies referred to as the ``GKV (gold)'' sample in Table~\ref{tab:sample}. The additional selections indicated in Table~\ref{tab:sample} are discussed further below in \S\,\ref{subsec:emu} and \S\,\ref{subsec:sample}.

\begin{figure}
\centering
\includegraphics[width=1\linewidth]{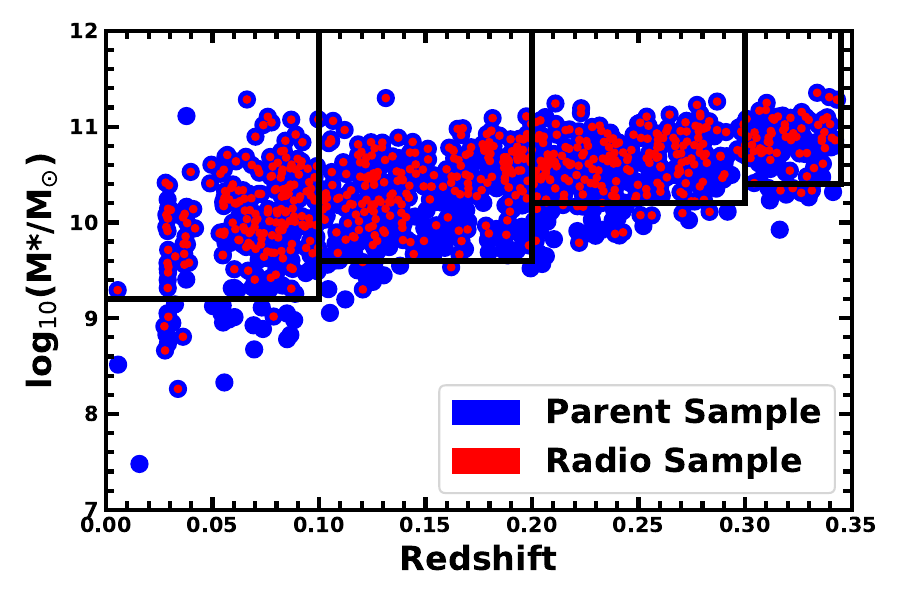}
\caption{Distribution of \textsc{ProSpect} $M_*$ of all galaxies in the sample as a function of redshift, illustrating the four volume limited samples with all the optical parent GKV galaxies (blue), and GKV - Radio detections (red). The black boundary lines are the redshift-dependent $M_*$ cuts for each redshift bin, with an apparent trend of increasing mass with redshift.}
\label{fig:GKV_sample}
\end{figure}

\subsection{Evolutionary Map of the Universe}
\label{subsec:emu}

EMU \citep{2025PASA...42...71H} is a radio continuum sky survey providing the most sensitive radio view of the entire Southern Sky, with sensitivity of $\sigma = 20\,\mu$Jy beam$^{-1}$ and resolution of $\sim 15"$ FWHM. EMU expects to deliver a catalogue of about 20 million radio sources at $943\,$MHz \citep{2021PASA...38...46N}. This analysis uses the EMU early science observations of G23 \citep{2022MNRAS.512.6104G} combined with the optical spectroscopy from GAMA. The G23 field is an 82.7 deg$^2$ region, spanning right ascension $338.1^{\circ}<$RA$<351.9^{\circ}$, and declination $-35^{\circ}<\delta<-30^{\circ}$ with $\sigma \approx 0.038\,$mJy beam$^{-1}$, at $887.5$\,MHz \citep{2019PASA...36...24L,2022MNRAS.512.6104G}. The catalogue for these observations includes 55\,247 radio sources of which 39\,812 have ${\rm S/N}\ge5$ \citep{2022MNRAS.512.6104G}. 

In this analysis, we consider only the sources common to both GAMA and the EMU early science radio data in G23. A $5$\,arcsec matching radius with the EMU catalogue is adopted, following the success curves reported in \citet{2024PASA...41...21A}. This results in a catalogue of 7\,123 ``GKV–Radio'' galaxies, of which 6\,999 galaxies have reliable redshifts ($nQ \ge 3$) (Table~\ref{tab:sample}). Within this group, a ``GKV–Radio (gold)'' sub-sample of 1\,700 galaxies (Table~\ref{tab:sample}) is defined by applying the more rigorous spectroscopic quality criteria outlined in \S~\ref{subsec:gama}.
We also convert radio luminosities from the measured $887.5$\,MHz to a frequency of $1.4$\,GHz using a spectral index\footnote{We adopt the convention that spectral index, $\alpha$, links flux density, $S$, and frequency, $\nu$, through $S\propto\nu^{\alpha}$.} of $\alpha=-0.7$, in order that we can subsequently use the SFR calibrations appropriate to $1.4$\,GHz luminosities.
The adoption of $\alpha = -0.7$ is commonly used, as it is the typical spectral index for synchrotron radiation, which is appropriate for radio emission in star forming galaxies. If we adopted more extreme values of spectral index (ranging from $\alpha = -0.9$ to $\alpha = -0.5$), the inferred 1.4\,GHz radio luminosities would only increase or decrease by 10\%. Any errors in the adopted spectral index would appear as a small random scatter within the dataset, rather than a significant systematic.

\begin{figure}
\centering
\includegraphics[width=1\linewidth]{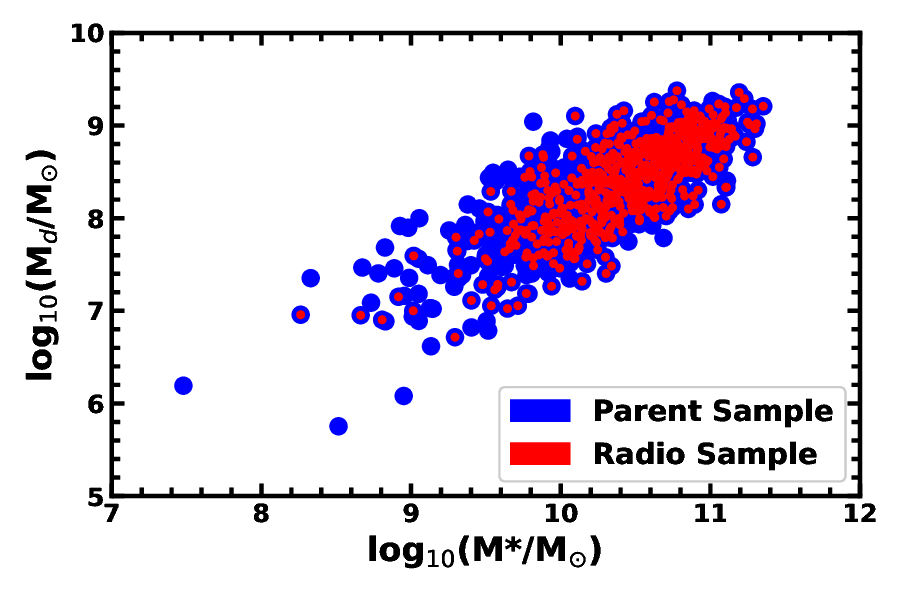}
\caption{\textsc{ProSpect} $M_d$ as a function of \textsc{ProSpect} $M_*$ for SFGs with $M_d$ having $S/N \geq 3$ in the optical parent GKV (gold) sample (blue), and the radio counterparts GKV-Radio (gold) (red). The radio detected sample has fewer low mass galaxies compared to the optical parent sample. These lower mass galaxies are missing the lower $M_d$ systems as well. The radio detected population favours higher mass galaxies compared to the optical parent sample, as detailed by \citet{2024PASA...41...21A}. The radio detected subset has a correlation coefficient of 0.79.}
\label{fig:MdMs_sample}
\end{figure}

\begin{table}
	\centering
	\caption{The $M_*$ lower limits, number of galaxies in GKV (gold), and GKV - Radio (gold), in each of the four mass-limited redshift bins.}
	\label{tab:zbins}
	\begin{tabular}{clcr} % four columns, alignment for each
		\hline
		\hline
		$z$ range & \(\log\left(\frac{M_*}{M_\odot}\right)\) & GKV & GKV - Radio\\
		  & & (gold) & (gold)\\
		\hline
		0 $<$ $z$ $\leq$ 0.1 & 9.2 & 697 & 231 \\
        0.1 $<$ $z$ $\leq$ 0.2 & 9.6 & 1332 & 302 \\
        0.2 $<$ $z$ $\leq$ 0.3  & 10.2 & 1042 & 252 \\
        0.3 $<$ $z$ $\leq$ 0.345 & 10.4 & 357 & 90 \\
		\hline
	\end{tabular}
\end{table}

\subsection{Final Sample}
\label{subsec:sample}

In this work, we choose the SED fitting tool \textsc{ProSpect} \citep{2020MNRAS.495..905R, 2020MNRAS.498.5581B} as a generative galaxy SED package, because \textsc{ProSpect} measurements are available as part of the GAMA G23 dataset. %We select \textsc{ProSpect} because it can model the SED all the way into the radio wavelengths.
\textsc{ProSpect} uses the BC03  \citep{2003MNRAS.344.1000B} and EMILES \citep{2016MNRAS.463.3409V} stellar population libraries. It uses a fixed Chabrier IMF \citep{2003PASP..115..763C}, the \citet{2000ApJ...539..718C} dust attenuation model, as well as the dust re-emission library of FIR templates by \citet{2014ApJ...784...83D}.
\textsc{ProSpect} has been applied to the GAMA data, fitting the photometry from the far-UV (FUV) to the FIR \citep{2020MNRAS.498.5581B}.
We draw on these existing measurements of the G23 galaxies to obtain the $M_*$ and the $M_d$ for our analysis. The performance of the \textsc{ProSpect} measurements has been analysed and quantified by \citet{2020MNRAS.495..905R, 2020MNRAS.498.5581B, 2021MNRAS.505..540T,2022MNRAS.517.6035T}.

\begin{figure}
\centering
\includegraphics[width=1\linewidth]{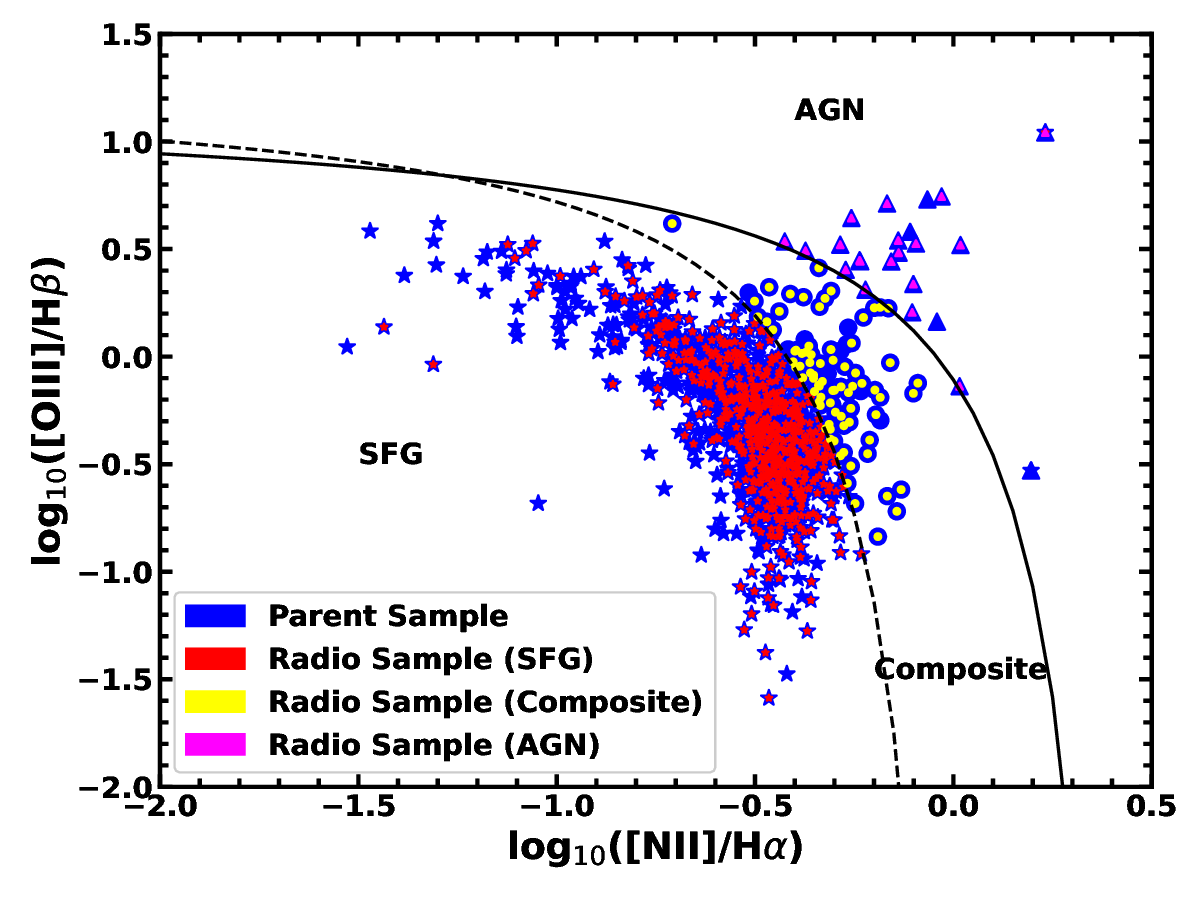}
\caption{Spectral diagnostic diagram illustrating the selection of star forming galaxies (SFGs), composite, and AGN populations. The selection boundaries are defined by the criteria given by \citet{2001ApJ...556..121K} (solid line) and \citet{2003MNRAS.346.1055K} (dashed line). Symbols represent the parent GKV sample in blue stars for SFGs, blue circles for composite objects, and blue triangles for AGNs. The GKV–Radio sample is highlighted in colour coded markers: red stars for SFGs, yellow circles for composite objects, and magenta triangles for AGNs.}
\label{fig:bpt}
\end{figure}

Figure~\ref{fig:GKV_sample} presents the \textsc{ProSpect} $M_*$ distribution of our sample as a function of redshift for the full GKV sample (blue), and radio counterparts (GKV - Radio, in red), which form the basis of our subsequent analysis. The full sample spans a redshift range $0 < z \leq 0.345$. The black boundary lines in Figure~\ref{fig:GKV_sample} represent the four independent volume limited samples. The $M_*$ range sampled by the radio counterparts lies toward the higher mass end of the galaxy mass distribution, an effect explored in detail by \citet{2024PASA...41...21A}. We use four volume limited samples for the bulk of our analysis, selected from those labelled as  ``GKV (gold)'' and ``GKV - Radio (gold)''. Constructing such samples enables an investigation of any potential evolutionary or mass dependent effects. For this analysis we construct our volume limited samples using a $M_*$ limit rather than the $r$-band absolute magnitude  limit adopted by \citet{2024PASA...41...21A}. We choose to use a $M_*$ limit here in order to cleanly analyse our results in terms of galaxy $M_*$. Our $M_{\ast}$ cut provides a physically consistent basis for comparing galaxies across different evolutionary stages. Choosing a sample based on $M_*$ instead of luminosity prevents our analysis from being biased by extreme dust obscuration that might otherwise hide such galaxies in an (observed) absolute magnitude limited sample. The redshift bins, $M_*$ limits imposed, and numbers of objects in each range are summarised in Table~\ref{tab:zbins}.

We define our final sample by retaining only those galaxies with \textsc{ProSpect} $M_d$ estimates having $S/N \geq 3$. From our GKV (gold) sample with 5039 SFGs, we are left with 1077 galaxies, and from the GKV-Radio (gold) sample we have 688 galaxies (Table~\ref{tab:sample}). Figure~\ref{fig:MdMs_sample} shows \textsc{ProSpect} $M_d$ as a function of \textsc{ProSpect} $M_*$ for these, with the optical parent sample (blue) overlaid with the radio counterparts (red). The radio detected subset is deficient in lower mass galaxies in comparison to the optical parent sample, as detailed by \citet{2024PASA...41...21A}, and accordingly is also missing the lower $M_d$ systems. The radio detected subset has a correlation coefficient of 0.79, while the optical parent sample has a correlation coefficient of 0.8 between ${M_d}$ and $M_*$. This will be important in our subsequent analysis where we choose to draw on a dust measurement that captures an underlying correlation with stellar mass.
\begin{figure}
\centering
\includegraphics[width=1.0\linewidth]{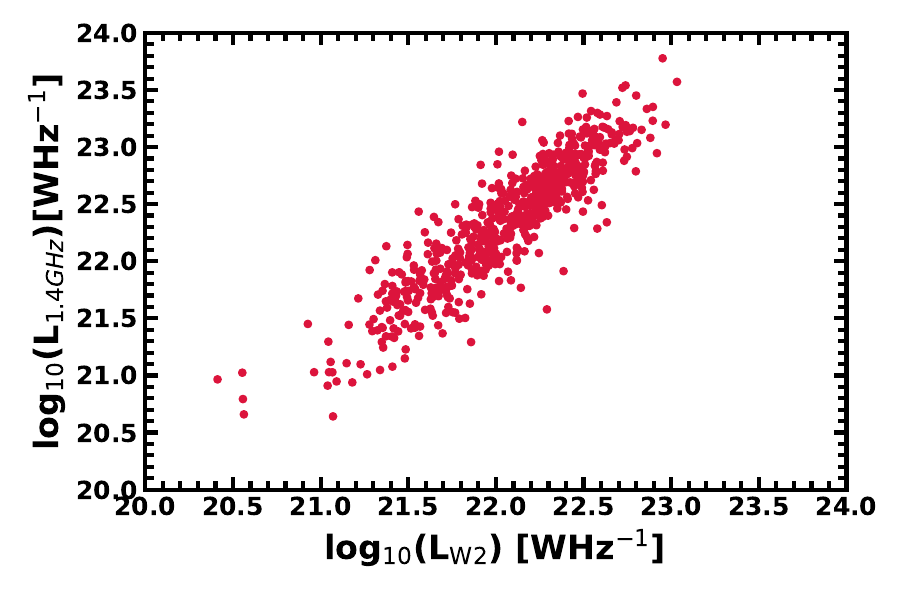}
\caption{The radio-IR correlation for our sample of radio-detected SFGs. The correlation between $L_{\rm 1.4GHz}$ and $L_{\rm W2}$ reflects the tight relationship expected for SFGs, with an absence of outliers above the main sequence. This lack of radio excess sources suggests that AGN activity does not contribute significantly to the measured radio luminosities.}
\label{fig:lum_WISE}
\end{figure}

Figure~\ref{fig:MdMs_sample} displays both the parent and the radio samples together, to illustrate the similarities and differences, as detailed in \citet{2024PASA...41...21A}. We do this as we rely only on the radio subsample in our remaining analysis because the radio measurements are required for our SFR calculations.
It has been noted \citep[e.g.,][]{2023MNRAS.523.3119W} that the $M_d$-$M_*$ relation may show increased dispersion for low stellar masses. This effect is not present in our current sample and will not affect any of our analysis, but it may be important to consider for samples at very low stellar mass.

Our initial sample contains both SFGs and AGN. Radio detections, in particular, often include AGN. As we aim to measure obscuration in SFG, we separate and exclude AGN dominated systems. Figure~\ref{fig:bpt} illustrates the standard diagnostic diagram of Baldwin, Phillips and Terlevich \citep[BPT;][]{1981PASP...93....5B, 1987ApJS...63..295V,2001ApJ...556..121K}, as detailed in \citet{2024PASA...41...21A} showing the optical parent sample and the radio counterparts. The data points in red stars, yellow circles and magenta triangles represent the radio SFGs, composites, and AGN, respectively, overlaid on the optical parent sample shown in blue (stars, circles, and triangles, respectively). The solid and dashed lines follow \citet{2001ApJ...556..121K} and \citet{2003MNRAS.346.1055K}. Galaxies below the dashed line are star forming. Those above the solid line are AGN. Systems that lie between the two are classified as composites. We keep these composite objects, but they are few and do not affect our conclusions.

To assess the effectiveness of excluding AGN using the BPT diagram,  we use optical spectra and redshifts from the IR–radio sample to investigate the radio–IR correlation. 
To identify the IR matches for our GKV parent sample, we used data from the ALLWISE catalogue \citep{2012yCat.2311....0C}. This catalogue comes from the WISE mission, which mapped the entire sky across four IR bands: W1 ($3.4\,\mu\text{m}$), W2 ($4.6\,\mu\text{m}$), W3 ($12\,\mu\text{m}$), and W4 ($22\,\mu\text{m}$). To identify the IR counterparts to our parent sample, we used positional cross-matching. If a WISE source was found within 3 arcseconds of a galaxy's location, we recorded it as a successful match.

Figure~\ref{fig:lum_WISE} shows the radio-IR correlation for our sample using the W2 luminosity $L_{\rm W2}$, and the 1.4 GHz radio flux density $L_{\rm 1.4GHz}$. The relationship is consistent with the tight correlation expected between $L_{\rm 1.4GHz}$ and $L_{\rm W2}$ for SFGs. Any strong radio AGN would appear as outliers above the correlation. This relation is effectively the same if $L_{\rm W4}$ is used instead of $L_{\rm W2}$, but the number of sources with W4 detections is dramatically smaller. We retain $L_{\rm W2}$ here to maximise the sample size. A systematic trend of AGN contamination in higher radio luminosity galaxies would lead to a tilt in the correlation, making it steeper than the one-to-one relation expected for star formation. Given that we do not see any such outliers, or such a tilt, this implies that there is no significant contribution to the radio luminosity by any AGN emission. Even if there is a low-luminosity AGN present that is contributing to the radio emission in a small number of galaxies, the dominant component of the radio emission will be associated with star formation.

We summarise the selection of our final sample under the following criteria:

\begin{enumerate}
 \item The redshift quality flag $nQ$ $\geq$ $3$,
 \item Signal-to-noise ($S/N$) limits on the H$\alpha$ and H$\beta$ Balmer lines of $S/N$ (H$\alpha$) $\geq 3$, and $S/N$ (H$\beta$) $\geq 3$,
 \item Positive detections of the [O{\sc iii}] and [N{\sc ii}] emission lines, for precise categorization of the galaxies into SF or AGN.
 \item Well-fitting of emission lines indicated by ${\rm NPEG}$ (H$\alpha$) $= 0$, and ${\rm NPEG}$ (H$\beta$) $= 0$ as line fitting flags,
 \item Redshifts between $0 < z \leq 0.345$ (Table \ref{tab:zbins}),
 \item Stellar masses above the limits given in Table \ref{tab:zbins},
 \item Classification as SFG using the BPT diagram (Figure~\ref{fig:bpt}).
\end{enumerate}

\begin{figure*}
\centering
\includegraphics[width=0.45\linewidth]{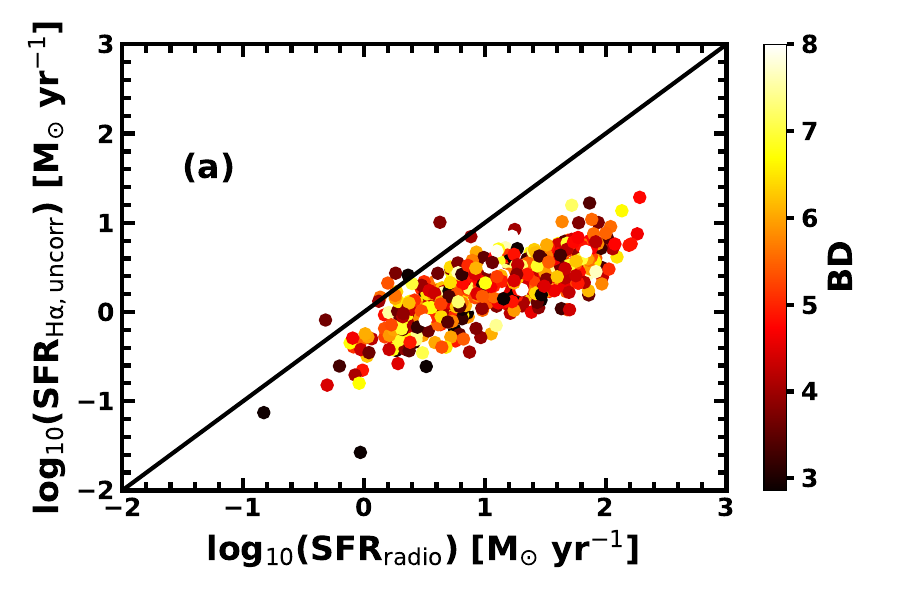}
\includegraphics[width=0.45\linewidth]{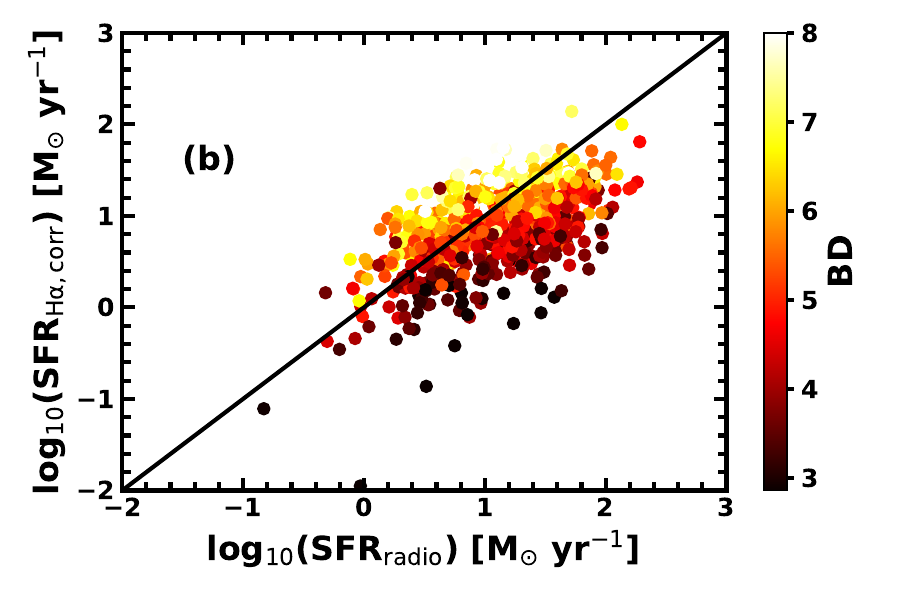}
\caption{Comparison between the SFR derived from $L_{\rm radio}$ and from uncorrected $L_{{\rm H}\alpha}$ in panel (a) and BD-corrected $L_{{\rm H}\alpha}$ in panel (b) for the radio sample, GKV-Radio (gold) colour coded by BD. The solid black line is the one-to-one line. The intrinsic scatter either side of the one-to-one line in panel (a) is $\approx 0.34\,$dex and panel (b) is $\approx 0.49\,$dex.  There is no BD gradient in the SFR comparison in panel (a) before applying the BD correction to $L_{{\rm H}\alpha}$. The correlation is tighter here, however, this correlation shows a deviation from the one-to-one line that increases with higher SFR, similar to observations in \citet{2001AJ....122..288H}. The trend in the BD in panel (b) is a result of applying the BD correction. BD generally increases with increasing $L_{{\rm H}\alpha}$ but decreases with higher $SFR_{\rm radio}$ at a fixed $SFR_{{\rm H}\alpha}$. This causes low BD points to shift slightly upward, while high BD points, originating from either end of the distribution, undergo a more significant shift.}
\label{fig:SFRHarad}
\end{figure*}

\section{RESULTS AND ANALYSIS}
\label{results_analysis}

\subsection{Star formation rates}
\label{subsec:SFR_Halpha_rad}

We are interested here in understanding the impact of dust on SFR estimates, and we investigate this using SFRs derived from H$\alpha$ and radio measurements \citep[e.g.,][]{2001AJ....122..288H, 2003ApJ...597..269A, 2003ApJ...599..971H, 2008MNRAS.386.1695S, 2016MNRAS.461..458D, 2018MNRAS.480..768D}. Part of this investigation is to understand the origins of the apparent SFR-dependent dust obscuration, identified differently by many authors using different samples and techniques \citep[e.g.,][]{2001AJ....122..288H, 2001ApJ...558...72S, 2003ApJ...597..269A, 2024PASA...41...21A}.

In order to investigate how galaxy obscuration characteristics are linked to SFR, we adopt the dust quantification approach previously developed by \citet{2025PASA...42...67F}.
This approach combines the BD and the $M_d$,  motivated by attempting to quantify and account for the presence of dust which is optically thick to H$\alpha$ emission. This approach can be considered as a way to quantify the dust geometry and is detailed further below in \S\,\ref{subsec:DustG}.
We begin first with a comparison of SFRs estimated from $L_{{\rm H}\alpha}$ and $L_{\rm radio}$ and the dependence on BD.

Radio emission from galaxies is insensitive to dust obscuration, making $L_{\rm radio}$ an advantageous tool for estimating SFRs in galaxies, once the presence of emission from any active galactic nucleus is excluded.

We first convert the $L_{\rm radio}$ and the $L_{{\rm H}\alpha}$ into SFRs using the common calibrations, reiterated briefly here. To calculate the SFR from H$\alpha$, we follow \citet{2003AJ....125..465H}. This begins with applying a stellar absorption correction to the H$\alpha$ equivalent width, $E{\rm W}_{{\rm H}\alpha}$: 

\begin{equation}
    S_{H\alpha}  =  \frac{E{\rm W}_{H\alpha}  +  E{\rm W}_c} {E{\rm W}_{H\alpha}} \times F_{H\alpha},
	\label{eq:S_hopkins}
\end{equation}

where $S_{\text{H}\alpha}$ is the stellar absorption–corrected line flux for H$\alpha$ and $F_{{\rm H}\alpha}$ is the observed line flux. $E{\rm W}_c$ denotes the Balmer line equivalent width correction for stellar absorption \citep[$E{\rm W}_c =2.5\,$\AA, following][]{2013MNRAS.433.2764G,2013MNRAS.430.2047H}. An equivalent relation is used to correct the H$\beta$ flux, using the same equivalent width correction.
The BD is calculated from these corrected line fluxes, BD $= S_{{\rm H}\alpha} / S_{{\rm H}\beta}$, where the nominal Case B value of BD = 2.86 \citep{1971MNRAS.153..471B, 1989agna.book.....O} indicates no obscuration.

In estimating SFR \citep{2003AJ....125..465H} from the ${{\rm H}\alpha}$ luminosity we adopt the calibration of \citet{1998ApJ...498..541K}:

\begin{equation}
    SFR_{H\alpha} [M_{\odot} yr^{-1}] =  \frac{L_{H\alpha}}{1.27 \times 10^{34} {\rm W}}, 
	\label{eq:sfr_hopkins}
\end{equation}

\noindent where $L_{{\rm H}\alpha}$, the luminosity of the H$\alpha$ emission line corrected for stellar absorption and dust obscuration is given by

\begin{multline} 
    L_{H\alpha} {(\rm W)} = (E{\rm W}_{H\alpha}  +  E{\rm W}_c)  \times  10^{-0.4(M_r - 34.10)}\\
    \times \frac{3 \times 10^{18}}{[6564.61(1+z)]^2}\times \left(\frac{\rm BD}{2.86}\right)^{2.36}
\label{eq:LHa}
\end{multline}

\noindent where $M_r$ is the $r$-band absolute magnitude \citep{2003AJ....125..465H, 2011MNRAS.415.1647G}. This equation implicitly assumes foreground screen dust, the attenuation curve of \citep{calzetti1994dust}, and that we can correct for this using the BD. 
This is the standard approach commonly used in making dust corrections using the BD.

Following \citet{2003ApJ...586..794B}, we adopt the calibration between 1.4\,GHz luminosity and SFR:

\begin{equation}
    SFR_{1.4GHz} [M_{\odot} yr^{-1}] =  \frac{f L_{1.4GHz}}{1.81 \times 10^{21} {{\rm WHz}^{-1}}},
	\label{eq:sfr_rad}
\end{equation}
where
\begin{equation}
f = 
\left\{
\begin{array}{ll}
 1 & L_{\rm 1.4GHz} > L_c \\
 (0.1 + 0.9 (L_{\rm 1.4GHz}/L_c)^{0.3})^{-1} & L_{\rm 1.4GHz} \le L_c,
\end{array}
\right.
\end{equation}
and $L_c$ = $6.4 \times 10^{21} \, \text{W\,Hz}^{-1}$. Adopting different calibrations \citep[e.g.,][]{2016MNRAS.461..458D, 2017ApJ...847..136B} for the $SFR_{\rm radio}$ does not substantially change our findings. While there would be small quantitative differences, the  qualitative results remain the same.

Both of these calibrations assume a Salpeter IMF. While choosing an alternative IMF would result in different scalings, they would make the same change to both equations~\ref{eq:sfr_hopkins} and \ref{eq:sfr_rad}. Consequently, comparisons between the SFRs using ratios of these quantities should be unaffected by the particular IMF chosen.

Figure~\ref{fig:SFRHarad}a illustrates the comparison between the SFRs estimated from the uncorrected $L_{{\rm H}\alpha}$ and $L_{\rm radio}$, while Figure~\ref{fig:SFRHarad}b presents the comparison between the SFRs estimated from the $L_{{\rm H}\alpha}$ corrected for dust attenuation using the BD and $L_{\rm radio}$. Both the figures are colour coded by BD, and the one-to-one line is represented by the solid black line. 
The apparent excess in the radio SFR compared to that inferred from H$\alpha$ is not a consequence of any AGN contribution to the radio luminosities, as evidenced by the tight radio-IR correlation shown in Figure~\ref{fig:lum_WISE}. This effect is strongly consistent with the infrared excess seen in Figure~5 of \citet{2009ApJ...703.1672K}, which compares $24\mu$m luminosities with H$\alpha$.

This effect is also not driven by bursty or stochastic star formation histories. \citet{2013MNRAS.434..209B} demonstrate that irregular recent starbursts are only significant for $M_*<10^{10} M_{\odot}$ galaxies. Our sample contains only 22\% of galaxies in that mass range, but this is also the range where the H$\alpha$ and radio SFR estimates are most closely aligned (Figure~\ref{fig:SFRHarad}).  
This suggests that bursty star formation histories are not a significant factor. 

Another potential explanation for a radio excess compared to H$\alpha$ in the SFRs is due to a timescale delay coming from radio emission associated with SF at somewhat earlier times compared to the current SF traced by H$\alpha$. This, too, cannot simply explain the offset between the H$\alpha$ and radio SFRs in Figure~\ref{fig:SFRHarad}. If that were the explanation, it would imply a highly unlikely coincidence between radio luminosity and star formation history. It would mean that the highest radio luminosity systems (those showing the greatest excess relative to H$\alpha$) are all tracing an equally large excess of recent star formation, rather than (as might be expected) a significant variation. The latter should show up instead as an increased scatter around a one-to-one relation, not a systematic offset.

In Figure~\ref{fig:SFRHarad}a, it is apparent that there is no gradient of BD in the SFR comparison before the BD correction to $L_{{\rm H}\alpha}$. The correlation is tight having a standard deviation orthogonal to the one-to-one line of $\approx 0.34\,$dex. The correlation here exhibits a deviation from the one-to-one line that increases with higher SFR \citep[as also observed in][]{2001AJ....122..288H}. This deviation is important for our analysis, and it varies depending on the BD correction applied. In addition, without any BD correction applied in Figure~\ref{fig:SFRHarad}a, it is evident that at higher luminosities, the suppression of H$\alpha$ relative to radio is more pronounced compared to lower luminosities.
We include the BD along with the $M_d$ because it provides insight into the underlying cause of the tilt in this relationship, which cannot be fully resolved by solely applying a BD correction. We could reproduce the empirical approach from \citet{2001AJ....122..288H} to make such a correction, however, our method (\S\,\ref{subsec:Hdust_SFR}) aims to explain it by incorporating the dust properties.

In Figure~\ref{fig:SFRHarad}b, the BD-dependent gradient is a consequence of applying the BD correction. It is apparent that there is a broad agreement between the two SFRs, although with the substantial intrinsic scatter that has been long recognised. We measure the scatter for this sample as having a standard deviation orthogonal to the one-to-one line of $\approx 0.49\,$dex. 
There are an equal number of high BD points distributed across the bottom and top ends of the distribution, scattered throughout. The gradient emerges after applying the correction because higher BD values correspond to higher SFRs, resulting in the observed gradient. While the general trend is evident, as expected, the $SFR_{\rm radio}$ tend to exceed those from the H$\alpha$, particularly at higher luminosities or SFRs \citep{1998ApJ...507..155C, 2001AJ....122..288H, 2001ApJ...558...72S}.

There is also a clear and strong trend visible with the BD. In general BD increases with increasing $L_{{\rm H}\alpha}$, but there is also a dependence on the $SFR_{\rm radio}$. At a fixed $SFR_{{\rm H}\alpha}$, for larger $SFR_{\rm radio}$ estimates the BD becomes {\em lower}. Additionally, the low BD points are slightly shifted upward, while the high BD points may begin at either end of the distribution, experiencing a more substantial shift. The BD correction contributes to the emergence of this gradient across the distribution. It is not immediately apparent why this should be the case, and this puzzle leads directly to the detailed exploration in the remainder of this analysis.

\begin{figure*}
\centering
\includegraphics[width=0.45\linewidth]{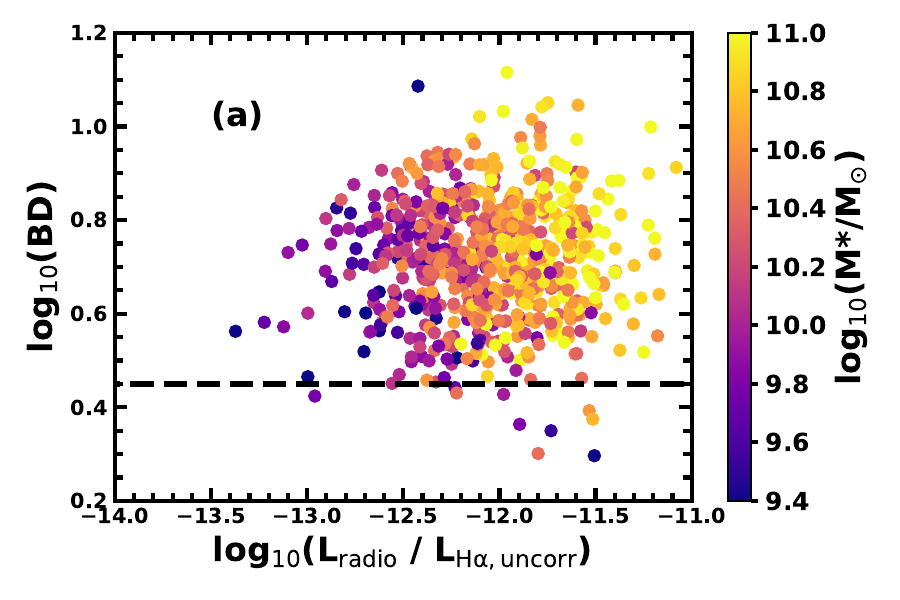}
\includegraphics[width=0.45\linewidth]{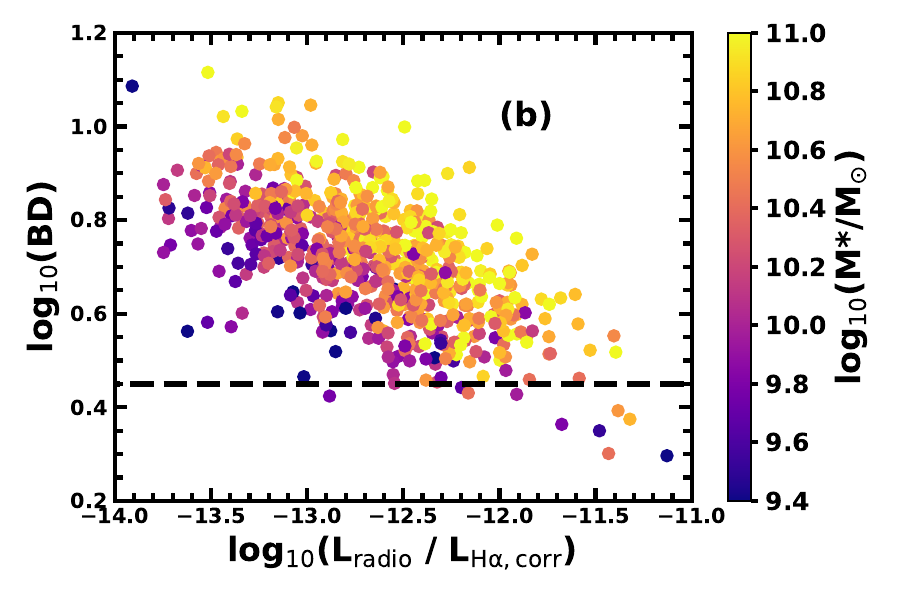}
\caption{The distribution of BD as a function of H$\alpha$ deficit (the ratio of the radio and H$\alpha$ luminosities) from uncorrected $L_{{\rm H}\alpha}$ in panel (a) and BD-corrected $L_{{\rm H}\alpha}$ in panel (b) colour coded by $M_*$. The horizontal dashed line represents the nominal Case B value of BD = 2.86. No clear relationship is apparent between the uncorrected H$\alpha$ and radio parameters and BD in panel (a). The tight correlation, in panel (b), is due to the high BD values significantly increasing the H$\alpha$, which shifts the H$\alpha$ deficit to the left, thereby inducing the correlation.}
\label{fig:BD_Hadeficit}
\end{figure*}

\begin{figure*}
\centering
\includegraphics[width=0.45\linewidth]{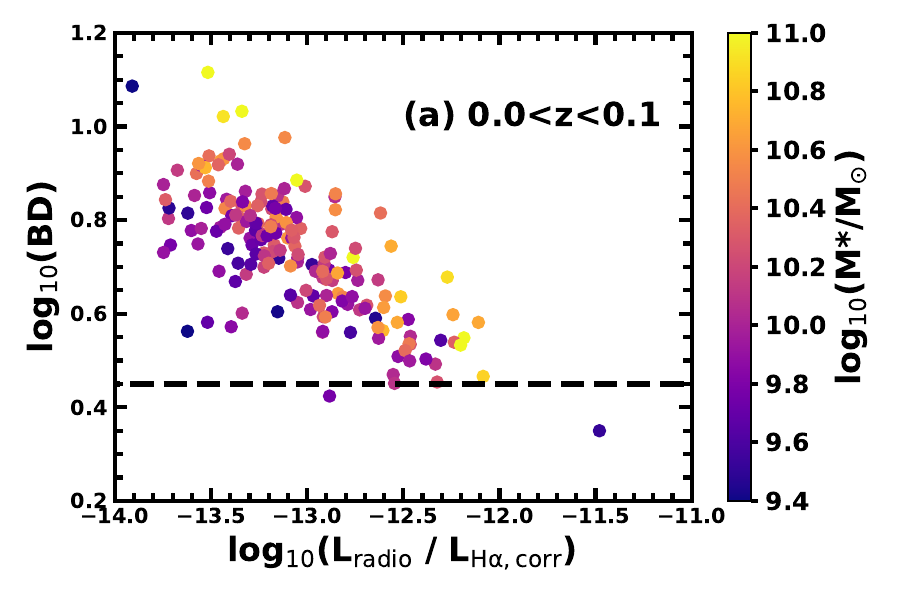}
\includegraphics[width=0.45\linewidth]{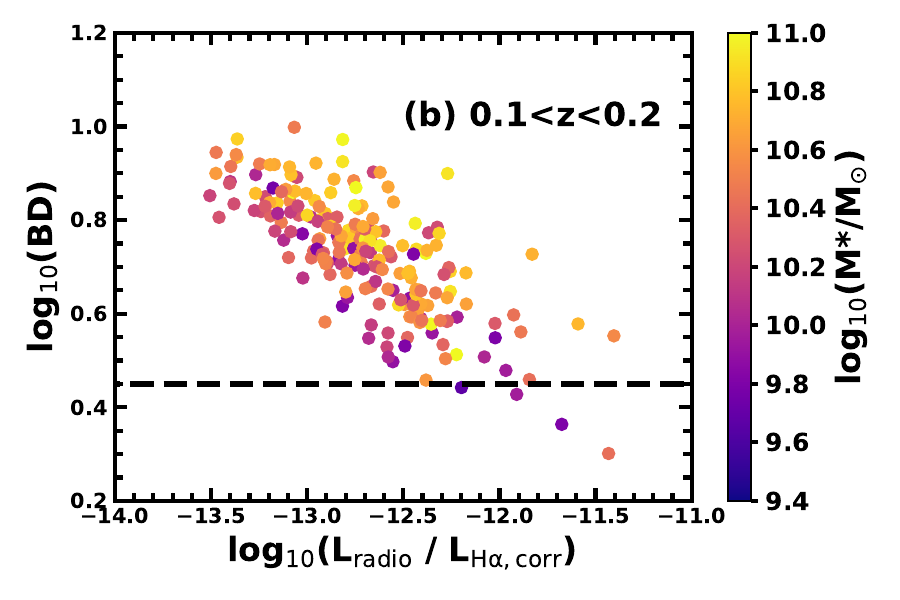}
\includegraphics[width=0.45\linewidth]{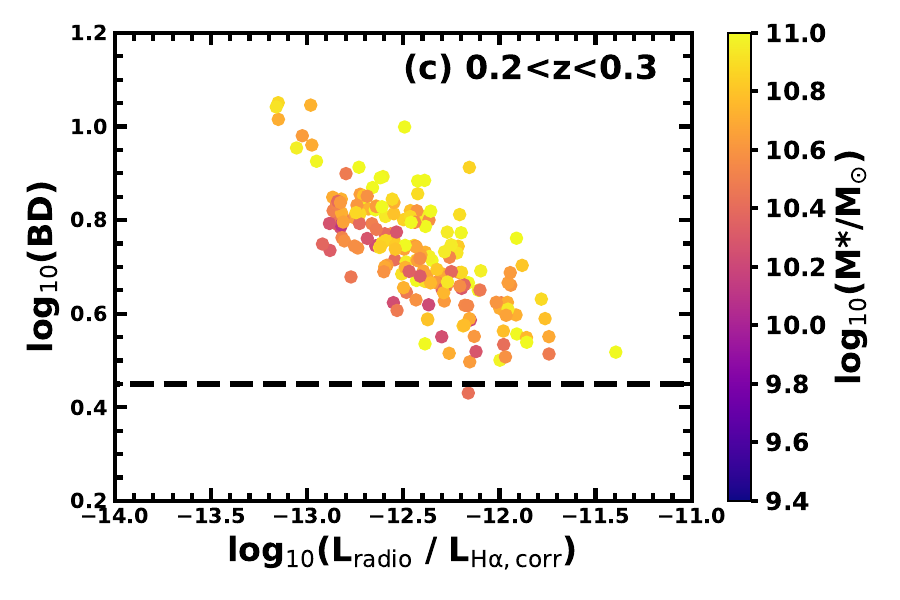}
\includegraphics[width=0.45\linewidth]{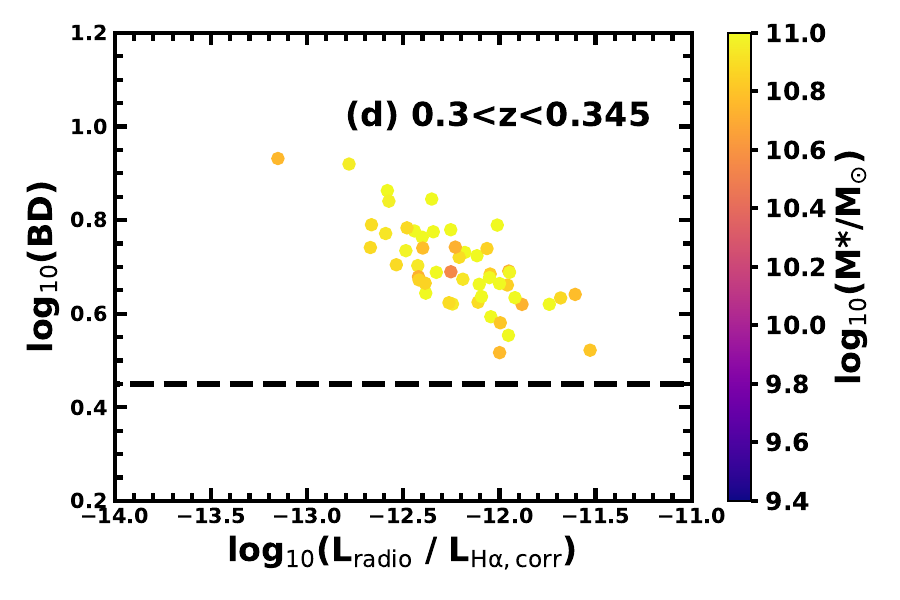}
\caption{The distribution of BD as a function of H$\alpha$ deficit colour coded by $M_*$ in four mass-limited samples. The horizontal dashed lines represent the nominal Case B value of BD = 2.86. In these mass-limited samples, higher redshifts are associated with subsets restricted to higher $M_*$ limits. BD dependence is present in each volume-limited sample, each probing a slightly different $M_*$, and with the increase in redshift, the trend shifts to the right, reflecting higher $M_*$ in higher redshift bins.} 
\label{fig:BD_Hadeficit_zbins}
\end{figure*}

Consider that galaxies with a large $SFR_{\rm radio}$  estimate compared to that from $SFR_{{\rm H}\alpha}$ may actually have an apparent ``radio excess'' by virtue of missing some H$\alpha$ emission due to its absorption by optically thick dust. If some of the H$\alpha$ emission associated with high-mass stars never escapes its host galaxy, it cannot be accounted for by making dust corrections to any H$\alpha$ that is actually observed. This leads us to draw on the concept of an ``H$\alpha$ deficit'', following \citet{2025PASA...42...67F},
by which we mean a relative underestimate in the H$\alpha$ derived SFR compared to that from an indicator insensitive to dust obscuration, such as $L_{\rm radio}$. 
We discuss these concepts in more detail in \S\,\ref{subsec:BD_Lrad}.

It may be the case that there are other physical explanations for the clear and luminosity-dependent deviation between $SFR_{{\rm H}\alpha}$ and $SFR_{\rm radio}$, such as variations in dust attenuation curves between galaxies, or the different timescales probed by the two indicators. It seems more likely that such effects would contribute a stochastic scatter within the relation than a strong and luminosity-dependent systematic. In any case, while we discuss the effect as arising from optically thick dust here, we are in effect simply quantifying an empirical relationship between the two SFR indicators. The results of our analysis are independent of the actual underlying physical cause. We refer the reader to \citet{2025PASA...42...67F} for further discussion of the origin of this effect.

\subsection{Radio SFR limitations}
\label{subsec:radioSFR_limitations}

The use of radio luminosities as SFR estimates has a long history \citep{1992ARA&A..30..575C}, and the limitations are well known. Clearly the biggest issue is to ensure that the galaxies are not hosting any AGN that may contribute to the radio emission. The BPT method is a reliable approach for identifying optical AGN, although it may not always be conclusive due to the absence of any of the four BPT lines
\citep[see, e.g.,][]{2010MNRAS.403.1036C}. When considering radio AGN \citep{2024PASA...41...16P}, it becomes more challenging, because weak accretion can produce radio emission without strongly ionising the ISM. Consequently, there are cases where galaxies classified as SF based on the BPT diagram may actually harbour radio AGN.  While such instances are relatively rare, they should be noted as potential contaminants \citep{2012MNRAS.421.1569B}. 

Excluding these AGN through standard diagnostics such as the BPT is perhaps the most common method, and this is the approach used here to ensure our sample consists of SFGs. We have used the radio-IR correlation (Figure~\ref{fig:lum_WISE} to confirm there are no AGNs with significant radio excesses present in our sample, that might otherwise explain an apparent H$\alpha$ deficit. While there is a possibility we have overlooked some low luminosity AGNs by solely relying on the BPT diagnostic, their inclusion in our sample would not significantly alter any of our findings.

Even for SFGs, though, there is a history of exploring systematics in the degree to which radio emission traces SFR. At low $M_*$, for example, a deviation from a simple linear scaling is seen \citep{1991ApJ...376...95C} and quantified differently by different authors \citep{2003ApJ...586..794B, 2017ApJ...847..136B}. We have adopted here the scaling by \citet{2003ApJ...586..794B}, although using that of \citet{2017ApJ...847..136B} changes our results only negligibly.
\citet{2018MNRAS.475.3010G} discuss alternative mechanisms for the generation of radio luminosity in low SFR galaxies, including pulsars or type Ia supernovae, as well as the role of enhanced magnetic fields in high-SFR galaxies, and their potential impact on radio SFR estimators.

While these and other limitations must be borne in mind, when we attempt to identify a reference SFR estimator that is independent of dust obscuration we are forced to the long wavelength regime, and the $L_{\rm FIR}$ and $L_{\rm radio}$. Despite their limitations, and fully anticipating that future work will go a long way toward improving their use as SFR probes, they currently provide the best starting point in estimating a total reference SFR in order to account for optically thick dust obscuration in galaxies.

\subsection{BD - luminosity ratio dependencies}
\label{subsec:BD_Lrad}

Here we explore the origins of the BD variations seen in Figure~\ref{fig:SFRHarad} between the two SFR indicators. To address this, we assess how BD correlates with the ratio of the two SFR measures. Rather than using SFRs explicitly though, we instead use the underlying luminosity measurements for simplicity. This eliminates the extra step of incorporating SFR calibration factors, keeping the analysis as close to direct observables as possible.
Throughout the manuscript, $L_{{\rm H}\alpha}$ and $SFR_{{\rm H}\alpha}$ refer to BD-corrected values unless stated otherwise.

Many authors have explored luminosity or flux ratios for similar purposes, either exploring links with obscuration \citep[e.g.,][]{2002AJ....124.3135K, 2006ApJ...637..242C, 2012MNRAS.426..330D, 2014MNRAS.441....2D, 2015MNRAS.453..879K}, or probing SFHs \citep[e.g.,][]{2019ApJ...881...71E, 2023MNRAS.526.1512R}. In almost all such cases it is the obscuration probed by a BD or similar optical attenuation parameter, such as E(B$-$V), being explored.

Motivated to explore the BD dependence seen in Figure~\ref{fig:SFRHarad}, Figure~\ref{fig:BD_Hadeficit} shows the distribution of BD as a function of the H$\alpha$ deficit (the ratio of the $L_{\rm radio}$ and the $L_{{\rm H}\alpha}$), colour coded by $M_*$. Figure~\ref{fig:BD_Hadeficit}a illustrates the distribution using uncorrected $L_{{\rm H}\alpha}$, while Figure~\ref{fig:BD_Hadeficit}b presents the distribution using the BD-corrected $L_{{\rm H}\alpha}$. 
The horizontal dashed line in both the figures represents the nominal Case B value of BD = 2.86. Throughout our analysis, the colour bar for $M_*$ spans the range \( 9.4 \leq \log\left( \frac{M_*}{M_\odot} \right) \leq 11 \) to provide a consistent reference for the $M_*$ range between figures. 

Although no clear relationship is evident in Figure~\ref{fig:BD_Hadeficit}a between the observed H$\alpha$ deficit and BD, applying a BD correction to the H$\alpha$ introduces a correlation (Figure~\ref{fig:BD_Hadeficit}b). This correlation arises due to the BD providing an insufficient correction to $L_{{\rm H}\alpha}$ in the presence of optically thick dust. This is also reflected in the fact that BD corrections alone do not fully correct the discrepancy between uncorrected H$\alpha$ and $SFR_{\rm radio}$ (Figure~\ref{fig:SFRHarad}). This underestimate is particularly relevant given the widespread practice of relying solely on BD corrections for inferring $L_{{\rm H}\alpha}$ and $SFR_{{\rm H}\alpha}$.
Our focus lies in resolving these discrepancies.

Looking in more detail at Figure~\ref{fig:BD_Hadeficit}b, in addition to the strong BD correlation with the H$\alpha$ deficit, there is an equally strong $M_*$ dependence, seen in the colour gradient. 
The higher H$\alpha$ deficit values occur for the lowest values of BD, but for a given BD, a higher H$\alpha$ deficit corresponds to higher $M_*$.
If optically thick dust is present, some H$\alpha$ emission cannot escape the galaxy to be measured, leading to an effective deficit in $L_{{\rm H}\alpha}$ compared to the $L_{\rm radio}$. This would correspond to data on the right in  Figure~\ref{fig:BD_Hadeficit}b which clearly show {\em low\/} values of BD. As explored by \citet{2025PASA...42...67F} we similarly interpret this as an indication that the measured H$\alpha$ in these systems arises only through the emission related to stars toward the outskirts of a ``distributed dust'' geometry, and hence the {\em detected\/} H$\alpha$ experiences little obscuration, although there may be significant H$\alpha$ that is fully obscured. Conversely, where the H$\alpha$ deficit is low, the BD is {\em high}. This would be consistent with a dust distribution arranged such that the emitted H$\alpha$ passes through the large proportion of it, producing the elevated BD, while suffering minimal optically thick losses, i.e., a ``foreground screen'' geometry. See \citet{2025PASA...42...67F} for more discussion of this effect.

We should note that our use of the phrases ``distributed dust'' and ``foreground screen'' are really just convenient descriptors of the dust distribution in the different regimes, and should not necessarily be taken too literally. The essential difference is that the former allows some of the SFR (and hence Balmer lines) to be totally blocked by dust and some to be visible through only slight extinction. The BD can therefore be small even though much of the H$\alpha$ is lost. The latter has dust moderately obscuring all the emission regions (not necessarily equally), so that BD can be larger yet a significant level of H$\alpha$ gets through.

Figure~\ref{fig:BD_Hadeficit_zbins} shows the same distributions in our four mass-limited samples spanning redshifts between $0 < z \leq 0.345$. This shows clearly that the BD dependence appears in each of the volume limited samples, each probing a somewhat different $M_*$ range. As redshift increases, the trend remains consistent, but it shifts to the right, a consequence of higher $M*$ values in progressively higher redshift bins.

It is apparent in Figure~\ref{fig:BD_Hadeficit} (also in Figure~\ref{fig:BD_Hadeficit_zbins}) that there are values of BD with high signal to noise line measurement that are still below the case B value. This occurrence in our sample is not a limitation of the measurements or the data; rather, it is common within similar galaxy samples. These measurements, although below the case B value, are reliable and not biased by artifacts in the spectra. 
The implication is that these galaxies probe slightly different physical conditions (temperature and electron density) from that assumed in case B. For these galaxies where the BD lies below the case B value, the physical conditions may differ, for example increased ISM temperatures would act to lower the BD values \citep{1989agna.book.....O}. Some of these values may, however, be too extreme to be explained in this way. If we simply chose to omit any measurements with BD$<2.86$, none of our analysis or conclusions would be altered. We retain them here for completeness, and further investigation into their detailed origin is deferred to future work.

\subsection{Dust Geometry Metrics}
\label{subsec:DustG}

\begin{figure}
\centering
\includegraphics[width=0.9\linewidth]{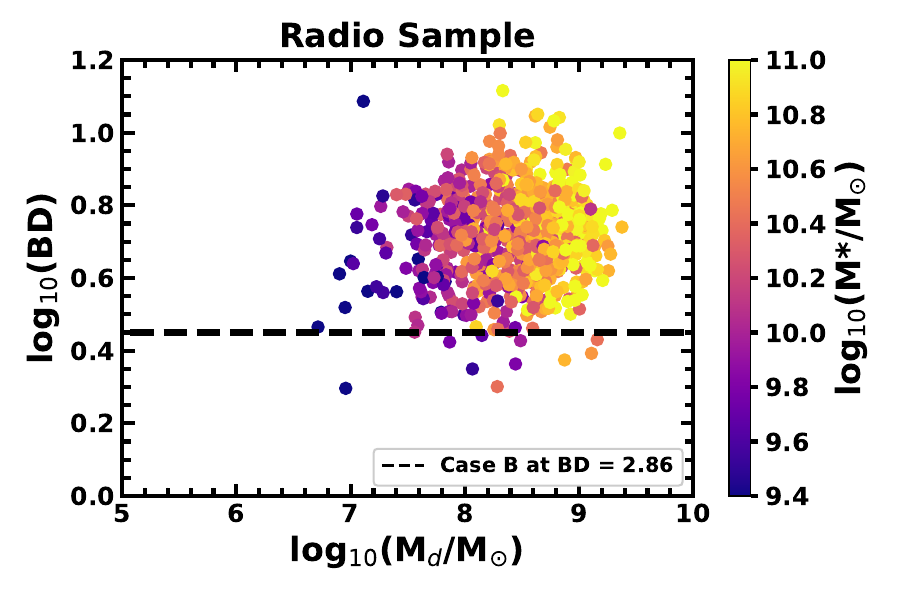}
\caption{The distribution of BD as a function of $M_d$ colour coded by $M_*$ for those with radio detection. The horizontal dashed line represents the nominal Case B value of BD = 2.86. The BD spans a wide range of values at the highest $M_d$, whereas only relatively low BD values are observed at low $M_d$, as shown also by \citet{2025PASA...42...67F}, who interpret this in terms of a dust geometry effect.}
\label{fig:BDMdMs}
\end{figure}

\begin{figure*}
\centering
\includegraphics[width=0.45\linewidth]{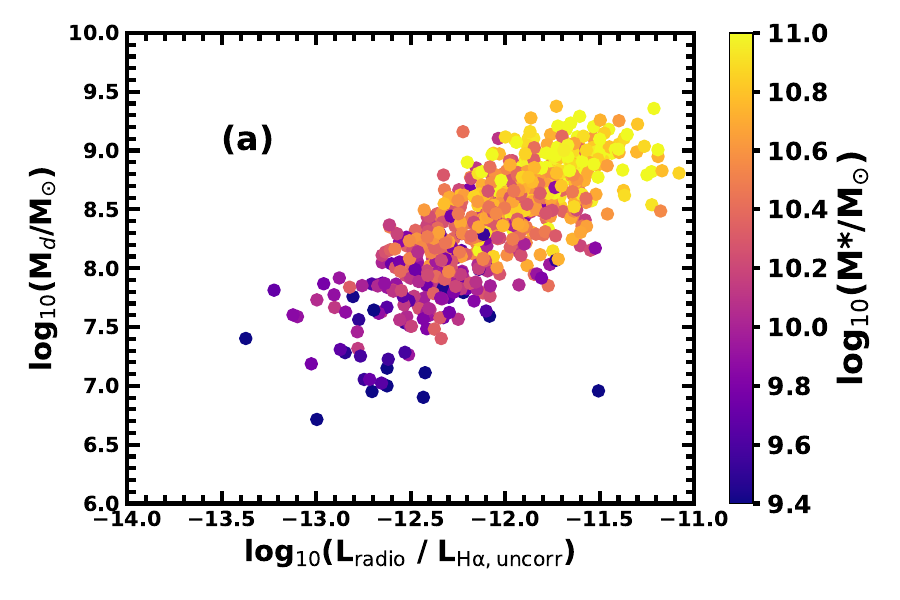}
\includegraphics[width=0.45\linewidth]{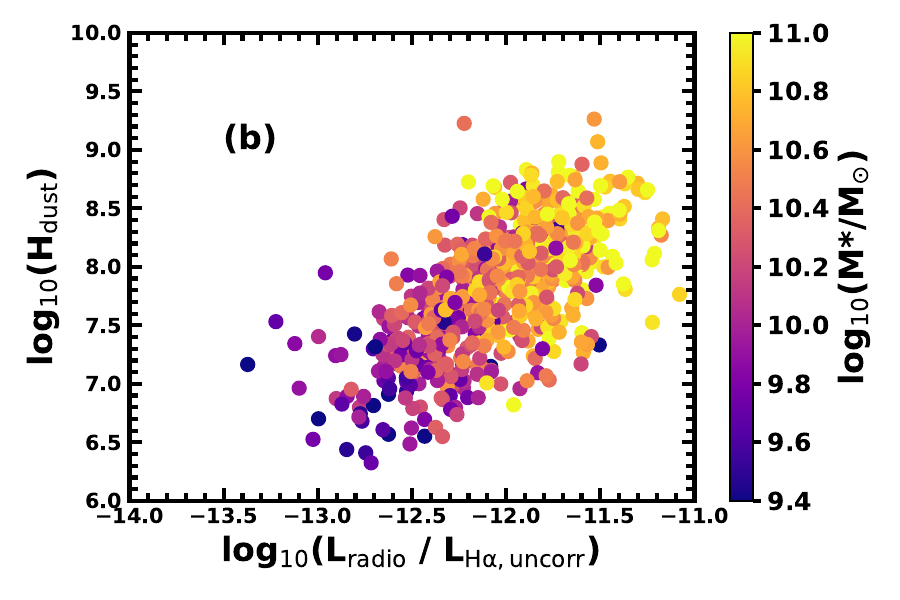}
\caption{The correlation between (a) $M_d$ and the H$\alpha$ deficit, and (b) $H_{\rm dust}$ and the H$\alpha$ deficit, using uncorrected $L_{H\alpha}$. These figures demonstrate that the observed correlations, such as that in Figure~\ref{fig:BD_Hadeficit}b, are not merely a consequence of both axes sharing a dependence on BD, as an equally strong correlation is seen here using the uncorrected $L_{H\alpha}$, which has no BD dependence.}
\label{fig:UncorrHa}
\end{figure*}

We are motivated to link BD and $M_d$ because relying solely on BD corrections does not fully rectify the attenuation of H$\alpha$ emission. The H$\alpha$ deficit is attributed to optically thick dust, for which the BD alone is insufficient to compensate. Likewise, the $M_d$ alone does not provide a straightforward means of correction, as it does not specify the extent of H$\alpha$ attenuation by the dust. Introducing the concept of dust geometry as used here, following  \citet{2025PASA...42...67F} in linking the BD and $M_d$, may enable a novel approach in accounting for optically thick dust. We test that idea here by using H$_{\rm dust}$ in estimating the SFR from the H$\alpha$ emission, to demonstrate, by comparison to the radio-derived SFR, that it accounts for the optically thick dust, in a way not possible by using the BD alone.

Following \citet{2025PASA...42...67F}, Figure~\ref{fig:BDMdMs} shows the distribution of BD as a function of $M_d$, for our analysis of the radio detected subsample.
The nominal Case B value of ${\rm BD} = 2.86$ is represented by the horizontal dashed line. This figure illustrates how galaxy obscuration spans the entire range of measurable BD at the highest $M_d$, but for lower $M_d$ is restricted to lower levels of BD. This effect is seen more clearly in \citet{2025PASA...42...67F}, where dust masses more than an order of magnitude smaller are being probed than in this sample.

As discussed by \citet{2025PASA...42...67F}, this relationship between BD and $M_d$ illustrates the distribution between the two extremes, interpreted there as (a) a maximal foreground screen geometry, if all the dust is arrayed in such a screen, and (b) a maximal distributed dust geometry, allowing for high optical depths. The low optical depth foreground screen geometry is situated at the lower end of the distribution, characterised by low BD and low $M_d$. Conversely, a high optical depth foreground screen geometry is located at the higher end of the envelope, characterised by high BD and high $M_d$. The absence of data in the upper left of the diagram is consistent with the idea that a sufficient amount of $M_d$ distributed entirely in a foreground screen would become totally optically thick to H$\alpha$.

Given that the BD correction alone is not sufficient to account for the deviation between the uncorrected $H\alpha$ and the radio SFR, (Figure~\ref{fig:SFRHarad}), we now draw on the approach of \citet{2025PASA...42...67F}, that links both the BD and $M_d$ in our aim to validate dust corrections.
Recall the strong BD dependence on $H\alpha$ deficit, which also shows a clear evidence of $M_*$ dependence (Figure~\ref{fig:BD_Hadeficit}). This implies that there is a parameter that relies on both BD and $M_*$ that should correlate strongly with the $H\alpha$ deficit. As in \citet{2025PASA...42...67F}, we construct such a parameter by using $M_d$ explicitly, due to its tight correlation with $M_*$, and linking it with the BD represented through the optical depth parameter $\tau$ \citep[e.g.,][]{calzetti1994dust}.
This is physically motivated by linking the $M_d$ (as an $M_*$ sensitive tracer) with the BD as an effective probe for sampling the dust properties, which should encapsulate both optically thick and thin dust corrections. We then calculate the same $H_{\rm dust}$ as in \citet{2025PASA...42...67F} :

\begin{equation}\label{eq:H_def}
    H_{\rm dust} = \frac{M_{d}}{10^{\tau^l_B}}.
\end{equation}
Here $M_d$ is the dust mass (in $M_{\odot}$).  We can also rewrite $H_{\rm dust}$
%by using Equation~\ref{eq:Tau_log}
to substitute an expression for BD in place of $\tau^l_B$ in Equation~\ref{eq:H_def} giving 

\begin{equation}\label{eq:H_BD}
   H_{\rm dust} = 10^{1.0578}\frac{M_{d}}{\rm{BD}^{2.303}}.
\end{equation}

The factor of 2.303 in the exponent of the BD accounts for the conversion from natural logarithms to base-10 logarithms, consistent with the formulation of extinction laws in terms of optical depths.

\begin{figure}
\centering
\includegraphics[width=1\linewidth]{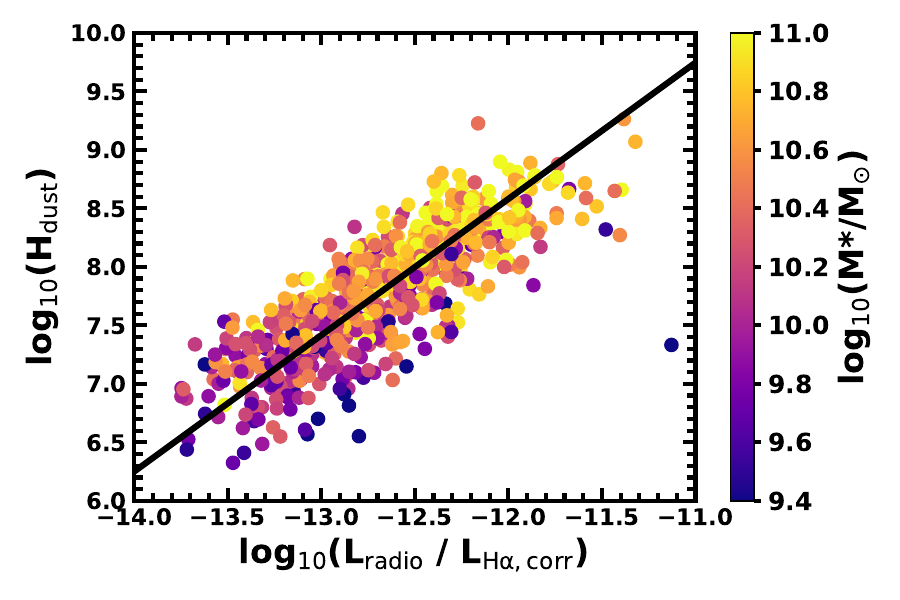}
\caption{The distribution of $H_{\rm dust}$ as a function of H$\alpha$ deficit, colour coded by $M_*$. The correlation between $H_{\rm dust}$, and H$\alpha$ deficit is significantly tighter compared to that of the correlation between BD and H$\alpha$ deficit in \S~\ref{subsec:BD_Lrad}. The correlation coefficient between $H_{\rm dust}$ and H$\alpha$ deficit is 0.81. The scatter compared to Figure~\ref{fig:BD_Hadeficit} is reduced to \(\approx 0.31 \, \text{dex}\). The solid black lines represent the lines fit as discussed in the text.}
\label{fig:Hdust_fit}
\end{figure}

We return now to the H$\alpha$ deficit, showing in Figure~\ref{fig:UncorrHa} and Figure~\ref{fig:Hdust_fit} how $H_{\rm dust}$ behaves.
To assess whether the correlation arises simply from a mutual dependence on BD, we consider the relationships between $M_d$ and the H$\alpha$ deficit, and $H_{\rm dust}$ and the H$\alpha$ deficit as shown in Figure~\ref{fig:UncorrHa}. In this case, we use H$\alpha$ values that are uncorrected for dust attenuation. In both comparisons, a strong correlation persists. This suggests that the trend is not solely an artifact of the BD dependence, but likely reflects an underlying physical connection.

\section{DISCUSSION}
\label{sec:discussion}

\subsection{Using \texorpdfstring{$H_{\text{dust}}$}{H dust} to improve \texorpdfstring{$SFR_{H\alpha}$}{SFR H alpha} estimates}
\label{subsec:Hdust_SFR}

The very tight correlation seen in Figure~\ref{fig:Hdust_fit} for $H_{\rm dust}$ (a scatter now of $\approx 0.31\,\text{dex}$, compared to $\approx 0.55\,\text{dex}$ in Figure~\ref{fig:BD_Hadeficit}) emphasises that $H_{\rm dust}$ is directly associated with the optical depth of the dust, corresponding to the H$\alpha$ deficit.
The correlation seen here is not a consequence of both axes sharing a dependence on BD, since there is an equally strong correlation seen between $M_d$ and the H$\alpha$ deficit calculated before making BD corrections to H$\alpha$ (Figure~\ref{fig:UncorrHa}).
The $M_*$ gradient orthogonal to the trend seen in Figure~\ref{fig:BD_Hadeficit} is essentially removed, although there is also a residual $M_*$ dependence for $H_{\rm dust}$, in the sense that $M_*$ increases with increasing $H_{\rm dust}$. This arises from the use of $M_d$ in the definition of $H_{\rm dust}$ indirectly introducing this scaling with $M_*$. 
To reiterate, at a given value of $M_d$ larger values of $H_{\rm dust}$ correspond to the maximal distributed dust model and the most optically thick dust for H$\alpha$.

We model the relationships for $H_{\rm dust}$ with the H$\alpha$ deficit using orthogonal regression line fitting. The fitted line is shown in Figure~\ref{fig:Hdust_fit}, and is given by:
\begin{equation}
    \log_{10} H_{\rm dust} = 1.165 \times \log_{10} \frac {L_{1.4GHz}} {L_{H\alpha}, corr} + 22.564
        \label{eq:H_dust},
\end{equation}

\begin{figure*}
\centering
\includegraphics[width=0.45\linewidth]{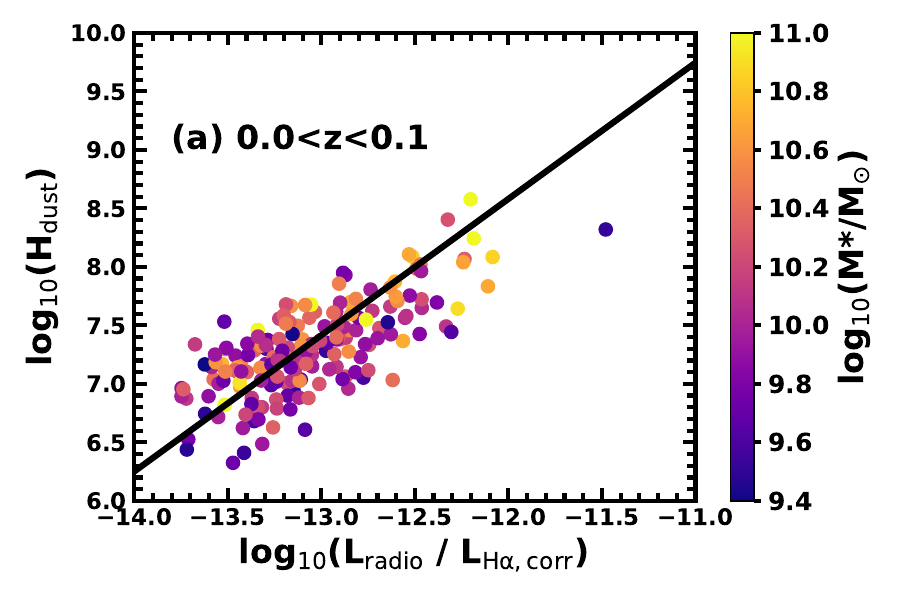}
\includegraphics[width=0.45\linewidth]{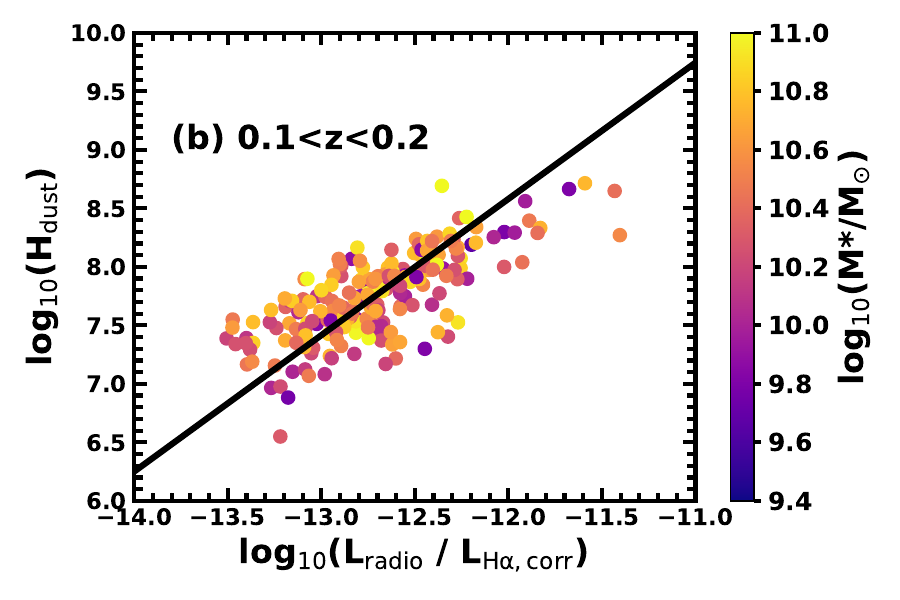}
\includegraphics[width=0.45\linewidth]{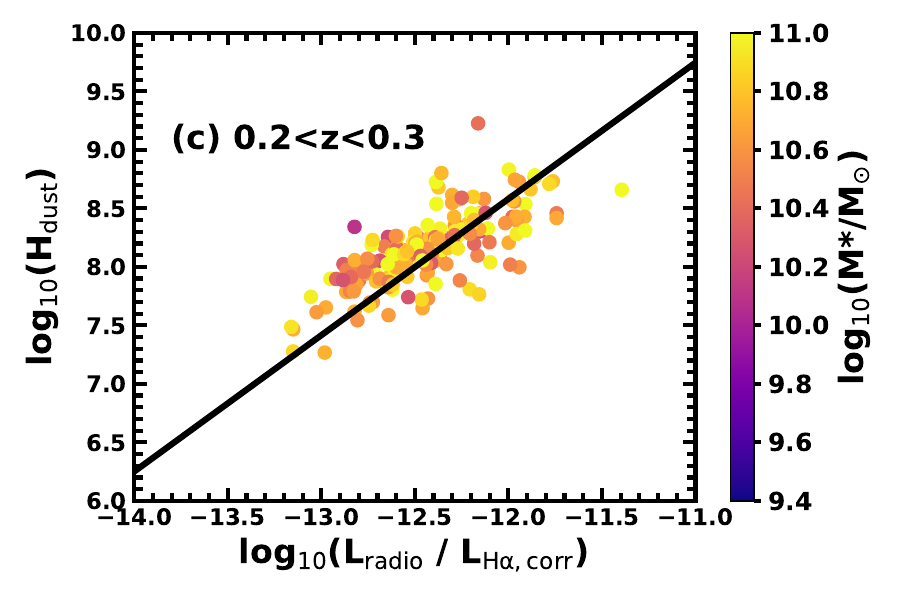}
\includegraphics[width=0.45\linewidth]{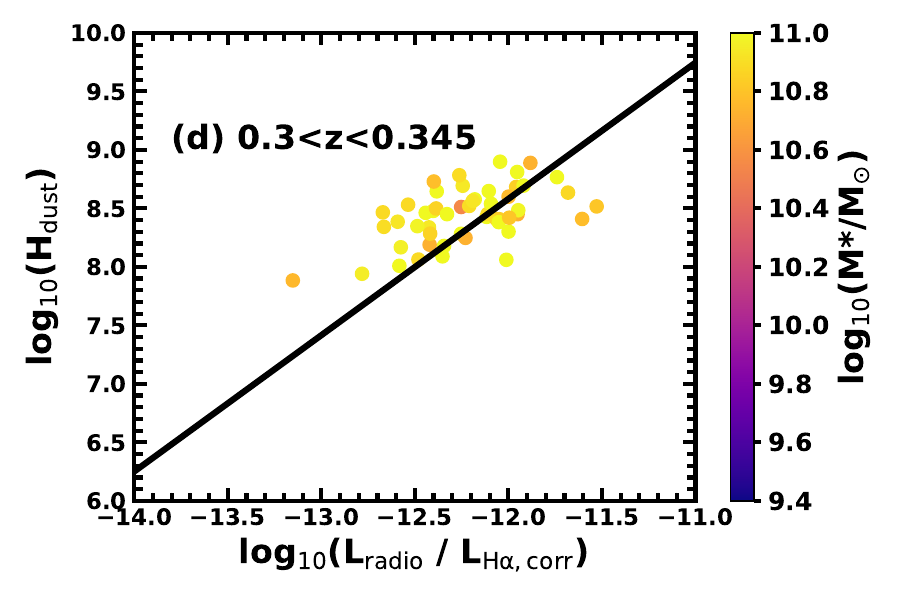}
\caption{The distribution of $H_{\rm dust}$ as a function of H$\alpha$ deficit, colour coded by $M_*$ in four mass-limited samples. The solid black lines in each panel represent the same line fit as in Figure~\ref{fig:Hdust_fit}. As redshift increases, the $M_*$ range sampled shifts to higher masses. However, there is no noticeable systematic change in the overall trend with redshift. This suggests that within the sampled redshift range, there is no redshift evolution in this correlation.}
\label{fig:Hdust_fit_z}
\end{figure*}

This same fit is reproduced in Figure~\ref{fig:Hdust_fit_z}, showing the mass-limited samples in each of the redshift bins. For $H_{\rm dust}$, while the $M_*$ ranges sampled in each redshift bin are reflected in the data distribution moving to higher masses at higher redshift, there is no apparent systematic shift in the overall trend as a function of redshift. This implies that there is no evolution in redshift for this correlation in the redshift range sampled here. 
For these reasons, we argue that the definition of $H_{\rm dust}$ is 
a self-consistent approach in defining a new parameter that is sensitive to the H$\alpha$ deficit, with minimal underlying $M_*$ and redshift dependencies.

\begin{figure*}
\centering
\includegraphics[width=0.45\linewidth]{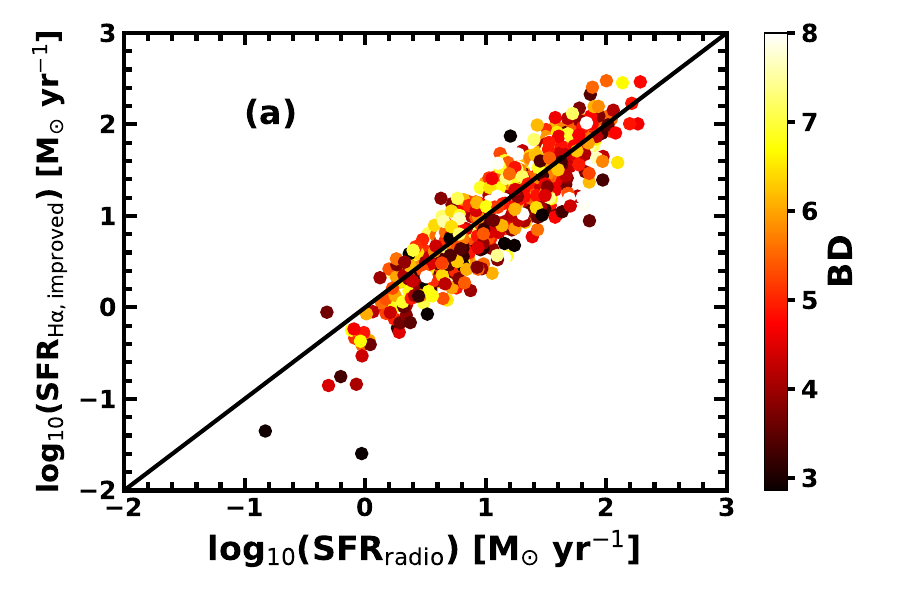}
\includegraphics[width=0.45\linewidth]{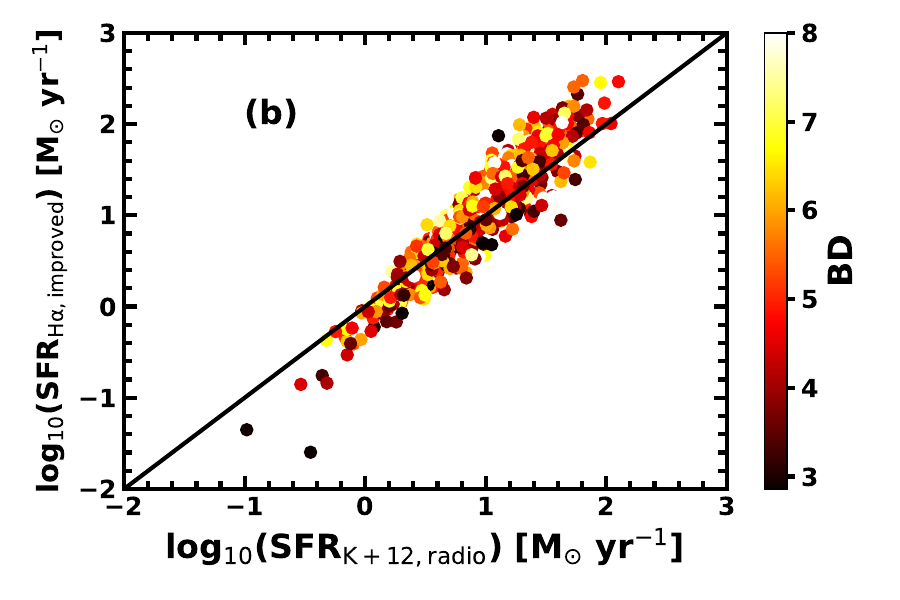}

\caption{Comparison between the updated H$\alpha$ SFR estimates from Equation~\ref{eq:Lum_Improved} together with Equation~\ref{eq:sfr_rad} and (a) the radio SFR; and (b) the composite SFR calibration from \citet{2012ARA&A..50..531K}. The solid black line represents the one-to-one relation and the measurements show a significantly tighter correlation, having a standard deviation in the scatter orthogonal to the one-to-one line of (a) $\approx~0.25\,\text{dex}$ and (b) $\approx~0.24\,\text{dex}$. The previously observed BD dependence is no longer apparent. The Pearson correlation coefficient in both the panels is 1.0.}
\label{fig:SFR_fit}
\end{figure*}

We are in a position now to use this new parameter to 
validate the SFR estimates from H$\alpha$, in essence to effectively correct for the emission otherwise lost to optically thick dust.

Early work by \citet{2001AJ....122..288H} recognised the discrepancy between SFRs obtained from UV and H$\alpha$ compared to those derived from FIR luminosities ($L_{\rm FIR}$) and $L_{\rm radio}$, which is more pronounced at high SFRs. The FIR and radio derived SFRs are unaffected by extinction, and the standard obscuration corrections to the UV and H$\alpha$ measurements are insufficient to fully address the discrepancies. They analyse the approach of incorporating a SFR dependent correction to these estimators. By introducing this empirical correction to a $SFR_{{\rm H}\alpha}$ estimate, in order to better link it to the $SFR_{\rm radio}$, they effectively accounted for H$\alpha$ otherwise missing due to obscuration by optically thick dust. Our approach follows the same philosophy by relying on the $SFR_{\rm radio}$ as a reflection of the full underlying SFR, but uses a more physically motivated approach to account for the optically thick dust affecting the H$\alpha$ estimators.

This relation can be rearranged as an expression for ${L_{\rm{1.4GHz}}}$, giving

\begin{equation}
    \log_{10} L_{1.4GHz}  = \frac {\log_{10} H_{\rm dust} - 22.564} {1.165} + {\log_{10} L_{H\alpha}, corr}
        \label{eq:Lum_Improved}
\end{equation}
If we use this expression with known values of $H_{\rm dust}$ and $L_{{\rm H}\alpha}$, we can estimate a predicted ``$L_{1.4GHz}$'', which may be considered to effectively be a refined estimate of $L_{{\rm H}\alpha}$, and hence $SFR_{{\rm H}\alpha}$, that now accounts for emission otherwise missing through obscuration by optically thick dust. 
This predicted $L_{1.4GHz}$ is then converted to a SFR using Equation~\ref{eq:sfr_rad}.

Figure~\ref{fig:SFR_fit}a shows the comparison of this newly defined $SFR_{{\rm H}\alpha}$ derived from Equation~\ref{eq:Lum_Improved} against the \, $SFR_{\rm radio}$, again colour coded by BD. This is contrasted against the original relation of Figure~\ref{fig:SFRHarad}. The solid black line represents the one-to-one relation between the SFRs, and the comparison now shows a significantly tighter correlation with a standard deviation in the scatter orthogonal to the one-to-one line of $\approx~0.25\,\text{dex}$. This is a significant improvement over the previous scatter of $\approx~0.49\,\text{dex}$. We evaluated the linear relationship using the Pearson correlation coefficient ($r = 1.00$) and and confirmed consistency with the Spearman rank correlation coefficient ($\rho > 0.999$). Notably, the previously observed BD dependence in Figure~\ref{fig:SFRHarad} is no longer apparent. 

It is worth noting that an alternative empirical approach to composite SFR calibration was proposed by \citet{2009ApJ...703.1672K}. Their approach is based on a linear combination of the observed $H\alpha$ and IR or radio luminosities. This combination effectively reproduces the true unattenuated $H\alpha$ luminosity.
We explicitly test our new method by comparing our results against the results from the composite SFR calibrations of \citet{2009ApJ...703.1672K}, (their equation 11 and Table 4), and \citet[][their Table 2]{2012ARA&A..50..531K}:

\begin{equation}
    SFR [M_{\odot} yr^{-1}] =  7.9 \times  10^{-35}[L_{H\alpha, uncorr} + 0.39 \times 10^{13} (L_{\rm 1.4GHz})]
\label{eq:sfr_kennicutt}
\end{equation}

\noindent where $L_{{\rm H}\alpha, uncorr}$ (in W) is the observed $H\alpha$ luminosity without correction for internal dust attenuation, ${\alpha}_{\lambda}$ = $0.39 \times 10^{13}$ is
the appropriate scaling coefficient \citep[from Table 2 of][]{2012ARA&A..50..531K} for the respective luminosity, for which we adopt $L_{\rm 1.4GHz}$, given here in units of $\text{W\,Hz}^{-1}$.

Figure~\ref{fig:SFR_fit}b shows the comparison of our newly defined $SFR_{{\rm H}\alpha}$ against this SFR calibrations by \citet{2012ARA&A..50..531K}, derived from Equations~\ref{eq:sfr_kennicutt}, colour coded by BD. The solid black line represents the one-to-one relation between the SFRs, and the results show a very tight correlation, with a correlation coefficient of 1.0. 
Our new method gives SFR estimates consistent with the approach of \citet{2012ARA&A..50..531K}. By adopting a physically motivated approach that incorporates $M_d$, we reproduce their empirical approach to combining $H\alpha$ and the radio data. Our method, however, is now driven directly by dust parameters and the $H\alpha$ only, so that it can be applied without explicit need for the radio or FIR data. Having $H\alpha$, $M_d$, and BD alone should be sufficient to estimate the SFR. While $M_d$ estimates themselves may in many cases originate from data that include FIR measurements, it may also be possible to draw on SED fitting from the UV and optical alone to derive $M_d$, where no FIR is available, although the uncertainties may be larger. The slight residual tilt observed in the relation is likely to be a consequence of our radio detected sample missing the low mass, least obscured objects at the low SFR end \citep{2024PASA...41...21A}.

This result demonstrates an approach to correcting the $SFR_{{\rm H}\alpha}$ estimates, which are now in much greater agreement, by construction, with the $SFR_{\rm radio}$. This refinement can be considered as a way to account, using the combination of BD and $M_d$ encoded within the $H_{\rm dust}$ parameter, for optically thick dust obscuration to H$\alpha$ that is otherwise not accessible. This method can be used to refine any $SFR_{{\rm H}\alpha}$ measurements, independent of the existence of any radio data, as long as both BD and $M_d$ are available. Since these data are typically accessible through optical spectroscopy and photometry, this approach can be applied to large H$\alpha$ samples regardless of radio detection.

Of course, this approach explicitly assumes that the $SFR_{\rm radio}$ estimate is a reliable tracer of the full underlying SFR in a galaxy, and our finding suggests an enhancement in the accuracy of the $SFR_{{\rm H}\alpha}$ estimates. This assumption has limitations, some of which are discussed below. But, while acknowledging these, the method presented here is put forward as a novel approach to improving the internal self-consistency of SFR indicators that allows for a way to account and correct for optically thick dust absorption. Based on our understanding of the $H_{\rm dust}$ parameter, this improvement serves to address optically thick dust obscuration affecting H$\alpha$, which would otherwise remain undetected.

\section{Conclusion}
\label{sec:concl}

In this analysis, we have validated the dust geometry metrics presented by \citet{2025PASA...42...67F}, that link BD and $M_d$, by demonstrating that they can be used in reliably correcting H$\alpha$ SFR estimates. Rather than replacing existing SFR estimation techniques, we have demonstrated that using the $H_{\rm dust}$ parameter can provide a physically complete representation of optically thick environments that traditional methods may overlook.

We start with a galaxy sample following \citet{2024PASA...41...21A} that provides the necessary optical and radio measurements. Using the dust geometry metric, $H_{\rm dust}$, we identify a clear link with the H$\alpha$ deficit, and use that directly to develop an improved $SFR_{{\rm H}\alpha}$ estimate (\S~\ref{subsec:Hdust_SFR}). Here we refer to the
ratio $L_{\rm radio}/L_{{\rm H}\alpha}$ as the H$\alpha$ deficit. 
We explore the effectiveness of drawing on the combination of BD and $M_d$ as quantified through the definition of $H_{\rm dust}$ in our analysis.
We demonstrate that using $H_{\rm dust}$ in this way proves to be a robust and effective approach in making obscuration corrections to H$\alpha$ SFRs.
Estimating $H_{\rm dust}$ simply requires a measurement of BD and $M_d$ (Equation~\ref{eq:H_BD}). The BD measurements are standard in most large spectroscopic surveys, and the dust mass may be estimated in a variety of ways, such as using SED modelling or direct measurement from FIR luminosities, depending on the sample.

We investigate the relationship between obscuration properties of radio detected galaxies and SFR in the local universe, using a sample of SFGs from the GAMA G23 region combined with the EMU early science data, following \citet{2024PASA...41...21A}.
The consistency between our results and independent radio and IR indicators confirms that this approach is effective for accurately estimating SFRs. This approach successfully captures the physical conditions of optically thick environments, making it a valuable addition to the standard tools used for estimating star formation.

Our main results can be summarised as:
\begin{enumerate}
    \item We use the novel dust geometry metric $H_{\rm dust}$ \citep{2025PASA...42...67F} to quantify the properties of dust in galaxies.
    \item We establish a significant and tight correlation between $H_{\rm dust}$ and H$\alpha$ deficit, which we interpret as due to the presence of optically thick dust.
    \item We used this tight correlation to define an estimate of $L_{{\rm H}\alpha}$, and hence $SFR_{{\rm H}\alpha}$ (Figure~\ref{fig:SFR_fit}). This reduces the scatter between the SFR estimators from an initial value of $\approx~0.49\,\text{dex}$ to $\approx~0.25\,\text{dex}$.
\end{enumerate}

If SFRs estimated from H$\alpha$ and UV are systematically underestimated, and more so at higher SFRs, that will have significant implications for many analyses. For example, SFRs at high redshift may need to be increased, perhaps by factors up to 5 or more (the effective underestimate in $SFR_{{\rm H}\alpha}$ compared to $SFR_{\rm radio}$ at high SFRs in Figure~\ref{fig:SFRHarad}). This will exacerbate the inconsistency between the cosmic SFR and mass density histories \citep{2006ApJ...651..142H,2008MNRAS.391..363W}. It will have broader implications in terms of understanding SFHs for individual galaxies, and will also impact many areas of high-mass SF and evolution. Future work will begin to explore these implications in more detail.
Such explorations will investigate the properties of dust in galaxies as a function of parameters including mass, SFR, redshift, galaxy type, and environmental factors. 
We plan to extend this work to resolved spectroscopy of galaxies using both current and upcoming integral field spectroscopy survey initiatives. These include surveys such as SAMI \citep{2012MNRAS.421..872C}, CALIFA \citep{2012A&A...538A...8S}, MANGA \citep{2015ApJ...798....7B}, MAGPI \citep{2021PASA...38...31F}, and TYPHOON \citep{2022ApJ...929..118G}.
This extension will enable us to gain a deeper understanding of the physical distribution of optically thick dust within galaxies.

\section*{Acknowledgements}

We thank the anonymous referee for important suggestions,
which helped improve the quality of this paper.

We thank Ian Smail, Alister Graham, and Tayyaba Zafar for helpful feedback and input as this work was being prepared. D.J.B. acknowledges funding from the German Science Foundation DFG, via the Collaborative Research Center SFB1491 ‘Cosmic Interacting Matters – From Source to Signal’.

GAMA is a joint European-Australasian project based around a spectroscopic campaign using the Anglo-Australian Telescope. The GAMA input catalogue is based on data taken from the Sloan Digital Sky Survey and the UKIRT Infrared Deep Sky Survey. Complementary imaging of the GAMA regions is being obtained by a number of independent survey programmes including GALEX MIS, VST KiDS, VISTA VIKING, WISE, Herschel-ATLAS, GMRT and ASKAP providing UV to radio coverage. GAMA is funded by the STFC (UK), the ARC (Australia), the AAO, and the participating institutions. The GAMA website is http://www.gama-survey.org/. Based on observations made with ESO Telescopes at the La Silla Paranal Observatory under programme IDs 179.A-2004 and 177.A-3016.

This scientific work uses data obtained from Inyarrimanha Ilgari Bundara / the Murchison Radio-astronomy Observatory. We acknowledge the Wajarri Yamaji People as the Traditional Owners and native title holders of the Observatory site. The Australian SKA Pathfinder is part of the Australia Telescope National Facility (https://ror.org/05qajvd42) which is managed by CSIRO. Operation of ASKAP is funded by the Australian Government with support from the National Collaborative Research Infrastructure Strategy. ASKAP uses the resources of the Pawsey Supercomputing Centre. Establishment of ASKAP, the Murchison Radio-astronomy Observatory and the Pawsey Supercomputing Centre are initiatives of the Australian Government, with support from the Government of Western Australia and the Science and Industry Endowment Fund.

%%%%%%%%%%%%%%%%%%%%%%%%%%%%%%%%%%%%%%%%%%%%%%%%%%
\section*{Data Availability}

All of the data are publicly available through GAMA and EMU. This paper includes archived data obtained through the CSIRO ASKAP Science Data Archive, CASDA (\url{http://data.csiro.au}).



\bibliographystyle{mnras}
\bibliography{dustsfr} % if your bibtex file is called example.bib

%\begin{thebibliography}{99}
%\end{thebibliography}

%%%%%%%%%%%%%%%%%%%%%%%%%%%%%%%%%%%%%%%%%%%%%%%%%%

%%%%%%%%%%%%%%%%%%%%%%%%%%%%%%%%%%%%%%%%%%%%%%%%%%

% Don't change these lines
\bsp	% typesetting comment
\label{lastpage}
\end{document}